\documentclass{llncs}
\usepackage[T1]{fontenc}
\usepackage{graphicx}
\usepackage{mathtools}
\usepackage[colorlinks=true]{hyperref}
\usepackage{color}

\usepackage{amsmath}
\usepackage{amsfonts}
\usepackage{amssymb}
\usepackage{xfrac}
\usepackage{graphicx}
\usepackage{multirow}
\usepackage{comment}
\usepackage[most]{tcolorbox}
\usepackage{cleveref}
\usepackage{bm}
\usepackage{comment}
\usepackage{booktabs}
\usepackage{todonotes}
\usepackage{algorithm}
\usepackage{algpseudocode}
\usepackage{adjustbox}
\usepackage{subcaption}
\usepackage{booktabs}
\usepackage{tabularx}

\DeclareMathOperator{\Z}{\mathbb{Z}}
\DeclareMathOperator{\R}{\mathcal{R}}
\DeclareMathOperator{\nums}{\nu_\mathsf{ms}}

\newcommand{\Vms}{V_\mathsf{ms}}

\newcommand{\Vs}{V_{s}}
\newcommand{\ams}{\frac{\delta_1}{p}}
\newcommand{\Vams}{V_{\ams}}
\def\Cov{\mathsf{Cov}}
\def\E{\mathbb{E}}
\def\V{\mathsf{Var}}
\def\nuenc{\nu_\mathsf{fresh}}
\def\nums{\nu_\mathsf{ms}}
\def\numul{\nu_\mathsf{mul}}
\def\U{\mathcal{U}}
\def\fc{\mathfrak{c}}   
\def\ks{\mathsf{ks}}
\def\ksk{\mathsf{ksk}}
\def\pk{\mathsf{pk}}
\def\sk{\mathsf{sk}}

\newtheorem{notation}{Notation}

\newenvironment{sketchofproof}
  {\begin{trivlist}
   \item[\hskip\labelsep\itshape Sketch of the proof.]}
  {\end{trivlist}}

  \def\ec{\color{black}}
\definecolor{rubblue}{RGB}{0, 53, 96}
\definecolor{rubgreen}{RGB}{141, 174, 16}
\definecolor{rubgray}{RGB}{231, 231, 231}

\newtcolorbox{algobox}[2][]{
    enhanced,
    breakable,
    sharp corners,
    boxrule=1pt,
    colback=white,
    colframe=rubblue,
    coltitle=rubblue,
    attach boxed title to top left={
        xshift=2mm,
        yshift=-2.5mm},
    boxed title style={
        tile,
        size=minimal,
        colback=white,
        left=1mm, right=1mm,
        before upper=\strut},
    title=#2,
    #1}

\begin{document}
\title{From Dependent Ciphertexts to Sharp Bounds for Polynomial Evaluation in BGV}
\titlerunning{Sharp Bounds for Polynomial Evaluation in BGV}
\author{Chiara~Marcolla\inst{1} \and%
    Nadir~Murru\inst{2} \and%
    Matilda~Urani\inst{3}
    }

\institute{%
    Technology Innovation Institute, Abu Dhabi, United Arab Emirates
    \and%
    Università degli studi di Trento, Trento, Italy
    \and%
    Politecnico di Torino, Torino, Italy}
\authorrunning{}
%
%
\maketitle             
\begin{abstract}
A central challenge in Fully Homomorphic Encryption (FHE) is the selection of ciphertext moduli, which depends on accurately estimating noise growth during homomorphic evaluation. Choosing excessively large moduli reduces the efficiency and scalability of the scheme, whereas underestimating noise can result in decryption failures and potential security vulnerabilities. Average-case analyses have provided promising probabilistic estimates of noise growth in practical FHE computations. However, recent studies indicate that reliable estimates must account for dependencies among ciphertext errors, as neglecting these correlations can systematically underestimate noise.

This work presents the first average-case analysis for the BGV scheme that accounts for both correlations induced by secret/public keys and ciphertext dependencies. We propose new variance estimates for the noise coefficients in homomorphic additions and multiplications involving dependent ciphertexts, and extend our framework to the homomorphic evaluation of polynomial functions, which is widely employed in FHE applications.  Our theoretical estimates closely align with experimental observations, exhibiting discrepancies smaller than $1.2$ bits. 

Finally, these estimates are adopted to choose the initial BGV parameters, leading to significantly smaller ciphertext moduli compared to those currently adopted in widely used FHE libraries, thus improving the efficiency of the scheme. Compared to OpenFHE, we obtain an average reduction of about $11.5\%$ in the ciphertext modulus size, while relative to HElib, we reduce the ratio between consecutive moduli up to 23 bits per level in practical settings.

\end{abstract}
%
%

\section{Introduction}
The increasing use of outsourced computation in settings such as cloud computing and machine learning raises fundamental concerns about data confidentiality.
Fully Homomorphic Encryption (FHE) addresses precisely this issue by enabling computations directly over encrypted data.
Since the first construction by Gentry in 2009 \cite{gentry2009fully}, an extensive body of research has led to several practical FHE schemes, including BGV \cite{brakerski2014leveled}, BFV \cite{brakerski2012fully,fan2012somewhat}, TFHE \cite{ducas2015fhew,chillotti2016faster,chillotti2020tfhe}, and CKKS \cite{cheon2017homomorphic,cheonRNSHEAAN18}.

Current FHE constructions base their security on the assumed hardness of the Learning with Errors (LWE) problem \cite{regev2005lattices}, and its ring variant (RLWE) \cite{lyubashevsky2010ideal}. 
Informally, the RLWE problem consists of distinguishing samples of the form $(a,b=s\cdot a+e)$ from uniformly random samples over the ring $\R_q=\Z_q[x]/\langle x^n+1\rangle$, where correctness in FHE requires the error term $e$ to remain sufficiently small compared to the ciphertext modulus $q$.

A fundamental challenge in FHE is therefore the selection of the ciphertext moduli.
Choosing unnecessarily large moduli significantly increases the computational cost and memory consumption of the scheme, reducing its practicality and scalability.
At the same time, moduli that are too small may compromise correctness or even security.
Since the noise grows during homomorphic evaluation, parameter selection critically depends on obtaining accurate estimates of the error growth throughout the computation \cite{marcolla2022survey}.

This issue is central to research in FHE and, over the years, various approaches have been proposed to address it. For example, employing the Euclidean norm \cite{brakerski2014leveled}, the infinity norm \cite{fan2012somewhat,kim2021revisiting}, and the canonical norm in \textit{worst-case} analysis \cite{costache2016ring,costache2020evaluating,digiusto2025breaking,gentry2012homomorphic,iliashenko2019optimisations,mono2022finding}. 

A promising approach in the recent literature is the adoption of the \textit{average-case analysis}, which models the noise coefficients as random variables and derives bounds according to the characteristics of their distributions. Interest in this method, initially applied in the TFHE scheme \cite{chillotti2016faster}, and subsequently in the CKKS \cite{costache2022precision,DBLP:journals/cic/BossuatCMNT25}, BGV \cite{murphycentral,costache2023optimisations,cryptoeprint:2025/2027} and BFV \cite{BFVvar} schemes, grew due to a recognized discrepancy between worst-case analyses and experimental behavior, as highlighted in \cite{costache2020evaluating}.
The introduction of the average-case approach offers a potential resolution to these disparities, as it makes possible to compute tight \textit{probabilistic} upper bounds. 

However, obtaining tight estimates is not sufficient.
For parameter selection, it is equally important to guarantee that the noise is never underestimated, even marginally.
Indeed, such underestimations can lead to two critical issues: first, ciphertexts may fail to decrypt correctly with non-negligible probability due to excessive noise; second, the scheme may become vulnerable to security attacks, as demonstrated in \cite{checri2024practical,cheon2024attacks}. 

Recent works on BGV \cite{cryptoeprint:2025/2027} and BFV \cite{BFVvar} have shown that, in order to avoid such underestimations of the noise, it is crucial to account for the dependencies among the errors associated with the different ciphertexts involved in a computation.
These dependencies arise because ciphertexts share common random components induced by the secret and public keys, making the corresponding noise terms inherently correlated.
These dependencies are intrinsic to any circuit evaluated on encrypted data, as all ciphertexts are generated under the same key.

However, this is not the only source of dependence among the error terms. In fact, in practical computations, additional forms of dependence naturally arise. To understand this, it is necessary to distinguish between \textit{common dependencies} and \textit{ciphertext dependencies} as discussed in \cite{cryptoeprint:2025/1036}. Common dependencies arise from the public and secret keys. Ciphertext dependencies, on the other hand, occur when the randomness involved in the encryption process is correlated, for instance, in operations involving the same ciphertext, such as self-products.
While for BFV \cite{BFVvar} and CKKS \cite{DBLP:journals/cic/BossuatCMNT25}, the study of specific circuits involving dependent ciphertexts has already been considered, only common dependencies are analyzed in the context of BGV.  In fact, the authors of \cite{cryptoeprint:2025/2027} focus only on the setting where the ciphertexts involved in the computation are mutually independent. However, in practical applications, homomorphic operations are frequently performed on dependent ciphertexts. Accurately accounting for such dependencies is therefore \textit{essential} to correctly estimate the noise growth and to avoid systematic underestimations of the resulting error.

Understanding these phenomena is becoming increasingly important in modern FHE applications, where homomorphic computations often involve the evaluation of high-degree polynomial functions and deep arithmetic circuits. This setting repeatedly combines dependent ciphertexts, naturally introducing correlations among the resulting error terms. As the multiplicative depth grows, worst-case analyses tend to produce increasingly conservative noise estimates, leading to unnecessarily large ciphertext moduli and reduced efficiency. 
Therefore, accurate methods to estimate the error growth are fundamental to selecting ciphertext moduli more efficiently.

\paragraph{Our contributions.}
In this work, we present the first BGV average-case analysis that accounts for both \textit{common dependencies}, induced by secret and public keys, and \textit{ciphertext dependencies}, arising from the use of correlated ciphertexts in homomorphic computations. 

First, we provide new estimates for the variance of the critical quantity after homomorphic additions (\Cref{prop.add.dep}), multiplications (\Cref{teo.mul.c1c2,teo.mul.c2}) involving potentially dependent ciphertexts, as well as after the product of a ciphertext and one of its rotations  (\Cref{thm:autoMS}).
Our analysis provides bounds that closely match experimental observations, with 
discrepancies in the $\log_2$ of the variance consistently below $1.2$ bits.

Second, we apply these techniques to the homomorphic evaluation of polynomials, where the same ciphertext is used many times and dependencies are therefore unavoidable. We derive noise estimates for two standard evaluation strategies, square-and-multiply (\Cref{teo.mul.ckpow.sm}) and baby-step 
giant-step (\Cref{teo.bsgs}). More generally, our framework applies to arbitrary polynomial evaluation strategies, once the corresponding sequence of homomorphic operations is specified. Indeed, the results developed in the previous theorems provide the basic building blocks needed to track the variance through additions, multiplications, and rotations of potentially dependent ciphertexts.
This allow us not only to estimate the resulting noise growth, but also to compare alternative evaluation methods and identify the most efficient one for a given polynomial.

Finally, we leverage these estimates to improve BGV ciphertext modulus selection. 
Compared to the parameter choices adopted in current FHE libraries, our methodology achieves smaller ciphertext moduli while preserving correctness guarantees.  In particular, when compared with OpenFHE \cite{OpenFHE}, our methodology reduces the ciphertext modulus by an average of approximately $16.3$ bits across all the evaluated benchmarks, corresponding to an average reduction of about $11.5\%$. Moreover, compared with the modulus ratios currently employed in HElib \cite{helib}, our approach reduces the size of the ratios between consecutive moduli by up to $23$ bits per level in practical parameter settings.\\

\paragraph{Roadmap of this work.} The paper is structured as follows. \Cref{sec:preliminaries}  introduces the notation and the probabilistic tools used throughout the work. \Cref{sec:BGV_scheme} gives an overview of the functioning of the BGV scheme. In \Cref{sec:theoric_contribution}  we provide accurate variance estimates of the critical quantity after the main homomorphic operations. Building upon these results, in \Cref{sec:poly-eval}, we explain how to correctly estimate the critical quantity resulting from the evaluation of generic polynomials. In \Cref{sec:experiments} we compare our estimates with experimental results and we show how the results of previous sections can be leveraged to correctly select the ciphertext moduli. We also compare our parameters with those currently adopted in widely used FHE libraries. Finally, \Cref{sec:conclusions} concludes this work by summarizing the main results.

\section{Preliminaries}\label{sec:preliminaries}
\subsection{Notation}
Let $n,t$, and $q$ be positive integers with $n$ a power of two.
We define $\R$ as the ring $\R =\mathbb{Z}[x]/\langle x^n+1\rangle$ and $\R_d = \mathbb{Z}_d[x]/\langle x^n+1\rangle$.
We denote by $M$ the multiplicative depth of the homomorphic circuit and by $L = M + 1$ the number of levels.  
We denote by $t$ the plaintext modulus, which must satisfy $t \equiv 1 \mod 2n$, from which the plaintext space $\R_t$ is constructed.
To support leveled homomorphic evaluation, we consider a chain of ciphertext moduli defined as
\[
q_\ell = \prod_{j=0}^{\ell} p_j, \quad  \, \ell \in  \{ 0,\dots,L-1\},
\]
where $q_\ell$ is the modulus used at level $L-1-\ell$. In particular, the initial modulus $q_{L-1}$ (i.e., at level $0$) will often be simply denoted by $q$.
In the following, polynomials are denoted by lowercase letters, while vectors of polynomials are represented in bold. For a polynomial $a$, we denote by $a|_i$ the coefficient of $x^i$. We use $[\cdot]_q$ to indicate the centered reduction modulo $q$ over the interval $[-q/2, q/2)$. When applied to a polynomial, the operation is intended to be coefficient-wise.

\subsection{Probabilistic distributions}
Given a polynomial $a \in \R$ and a probability distribution $\mathcal{X}$, the notation $a \leftarrow \mathcal{X}$ denotes that each coefficient of $a$ is sampled independently according to $\mathcal{X}$. 
We denote by (i) $\mathcal{U}_q$ the uniform distribution over $\mathbb{Z}_q$ with representatives in $\left[-q/2, q/2\right)$; 
(ii) $\mathcal{DG}_q(0,\sigma^2)$ the discrete Gaussian distribution, obtained by sampling from a Gaussian distribution with mean $0$ and variance $\sigma^2$, rounding to the nearest integer, and reducing modulo $q$, with representatives in $\left[-q/2, q/2\right)$.
Moreover, we denote by $V_a$ the variance of the coefficients of a polynomial $a$ and by $V(a|_i)$ the variance of the $i$--th coefficient. 
In particular, abusing the notation, we have that $V_{\mathcal{DG}_q(0,\sigma^2)} = \sigma^2$, $V_{\mathcal{U}_3} = {2}/{3}$, and $V_{\mathcal{U}_q} = {(q^2 - 1)}/{12} \approx {q^2}/{12}$.
Finally, we recall the following well-known results.

 \begin{proposition}[\cite{feller1950introduction}]\label{probfacts}
 Let $X,Y,Z,$ and $W$ be random variables. Then,
\begin{enumerate}
     \item if $X,Y$ are independent w.r.t. $Z,W$, $\Cov(X,Y)=0$ and $\E[X] = 0$, then $\Cov(XZ,YW)=0$.
    \item if $X,Y$ are independent and $\E[X] =\E[Y] = 0$, then $\V(XY)= \V(X)\V(Y).$
    \item if $X,Y$ are independent then $\V(X+Y)= \V(X)+\V(Y)$ otherwise $\V(X+Y)= \V(X)+\V(Y)+ 2\Cov(X,Y).$
    \item for any $\gamma \in \Z$, $\V(\gamma X)= \gamma^2\V(X).$
    \item if $X,Y$ have finite second moments, then $\left|\E[XY]\right| \le \sqrt{\E[X^2] \cdot \E[Y^2]}$ (Cauchy–Schwarz Inequality).
\end{enumerate}
\end{proposition}

\begin{lemma}[\cite{SteinShakarchi03}]\label{lem:parseval}
Let $n$ be a power of two and let $\zeta_1,\dots,\zeta_n$ be the $n$ roots of unity. Then, for every
$z = \sum_{j=0}^{n-1} z|_j\, x^j \in \R$,
\[
  \sum_{j=0}^{n-1} \left(z|_j\right)^2 \;=\; \frac{1}{n}\sum_{m=1}^{n} \left|z(\zeta_m)\right|^2 .
\]
In particular, if $z$ is a random polynomial whose coefficients have finite second moments,
\[
  \sum_{j=0}^{n-1} \mathbb{E}\!\left[(z|_j)^2\right]
  \;=\; \frac{1}{n}\sum_{m=1}^{n} \mathbb{E}\!\left[\left|z(\zeta_m)\right|^2\right].
\]
\end{lemma}
\ec 

\section{The BGV scheme}
\label{sec:BGV_scheme}
 In this section, we recall the three basic functions of the BGV scheme and the homomorphic operations of interest: addition, multiplication, modulus switching and rotation. 

\begin{algobox}{Key Generation.}
    Sample $s \leftarrow \chi_s$, $a \leftarrow \mathcal{U}_{q_{L-1}}$ and $e \leftarrow \chi_e$. \\ 
The key generation outputs the secret key $\sk = s$ and the public key $\pk = (b,a) = (-a\cdot s + t e, a)~\bmod~q_{L-1}.$
\end{algobox}
\newpage

\begin{algobox}{Encryption.}
Given the plaintext $m \in \R_t$ and the public key $\pk = (b,a)$, sample $u \leftarrow \chi_s$ and $e_0,e_1 \leftarrow \chi_e$. The encryption outputs $\fc = (\bm c, {q_{L-1}}, \nuenc)$, where 
\[
\bm{c} = (c_0,c_1) = (b u + t e_0 + m,\; a u + t e_1) \bmod q_{L-1},
\]
where $u,e_0,e_1 \in \R_{q_{L-1}}$ and $\bm{c} \in \R_{q_{L-1}}^2$.
\end{algobox}
\begin{algobox}{Decryption.}
    Given $\fc =(\bm{c}, q_\ell , \nu)$ and the secret key $\sk = s$.\\ 
    The decryption outputs $m = \left[ \left[ c_0 + c_1 \cdot s \right]_{q_\ell} \right]_t$. 
\end{algobox}
We adopt the term \textit{extended ciphertext} to refer to the tuple $\fc = (\bm{c}, q_\ell, \nu)$ where $\bm{c}$ is the ciphertext, $q_\ell$ is the corresponding modulus, and $\nu$ is the associated critical quantity. The \textit{critical quantity}  $\nu = [c_0 + c_1 \cdot s]_{q_{\ell}}$ of a ciphertext $\bm c$ defines whether $\bm c$ can be correctly decrypted. 
\noindent A proof of the correctness of the decryption process can be easily verified for a fresh ciphertext. In fact, the critical quantity can be rewritten as 
\begin{equation*}
    \nuenc = [c_0 + c_1 \cdot s]_{q_{L-1}} = [ m + t(e \cdot u + e_1 \cdot s + e_0) ]_{q_{L-1}} = [ m + t \varepsilon ]_{q_{L-1}},
\end{equation*}
where $\varepsilon$ denotes the \textit{error} introduced during encryption.
Since the critical quantity is simply the message perturbed by an error term whose coefficients are multiples of $t$, it is immediate to verify that reduction modulo $t$ recovers the original message if the error remains below a certain threshold. Precisely, to ensure the correctness of the decryption, it is necessary that \cite{mono2022finding}:    $\| \nu \|_{\infty}  < \frac{q_\ell}{2}.$
Otherwise, $m + t\varepsilon$ may wrap around the modulus, causing decryption to output a random polynomial rather than the original message.

When operations are performed on ciphertexts, they are reflected in the associated critical quantity, which grows with each operation \cite{costache2016ring}.
Therefore, to correctly select the ciphertext moduli, it is fundamental to accurately estimate the magnitude of this critical quantity, commonly referred to as the \textit{noise}.
\paragraph{Variance of the error coefficients.} Adopting an average-case perspective, we are thus interested in estimating the variance of the coefficients of the critical quantity $\nu$. Indeed, this variance allows us to derive tight probabilistic bounds on the magnitude of the noise and, consequently, to select ciphertext moduli that are as small as possible while still ensuring correct decryption.\\
For the critical quantity of a fresh ciphertext, we have that \cite{cryptoeprint:2025/2027}
\begin{equation}\label{eq.V.enc}
            V_\mathsf{clean} = \V(\nuenc|_i) = {t^2} \left(\tfrac{1}{12} + nV_eV_u + V_e + nV_eV_s \right).
\end{equation}

\begin{algobox}{Homomorphic addition.}
    Given two extended ciphertexts $(\bm{c}, q_\ell , \nu)$ and $(\bm{c}', q_\ell , \nu')$, the homomorphic addition outputs $(\bm{c}_{\mathsf{add}}, q_\ell , \nu_{\mathsf{add}})$, where
    \begin{equation*}
     \bm{c}_{\mathsf{add}}=(c_0 + c_0' , c_1 + c_1' ) \mod q_\ell,
\end{equation*}
with $\nu_{\mathsf{add}} =\nu + \nu'$.  
\end{algobox}

\noindent We are going to study the variance $\V(\nu_{\mathsf{add}}|_i)$ in \Cref{prop.add,prop.add.dep}.

\begin{algobox}{Homomorphic multiplication.}
    Given two extended ciphertexts $(\bm{c}, q_\ell , \nu)$ and $(\bm{c}', q_\ell , \nu')$, the homomorphic multiplication outputs $((d_0, d_1,d_2), q_\ell, \numul)$, where
    \begin{equation*}
    (d_0, d_1,d_2) = (c_0 \cdot c_0' , c_0 \cdot c_1' + c_1 \cdot c_0', c_1 \cdot c_1' ) \mod q_\ell,
\end{equation*}
with $\nu_{\mathsf{mul}} =\nu \cdot \nu' =  d_0 + d_1 s  + d_2 s^2$.  
\end{algobox}
\noindent It is worth noting that the product of two ciphertexts (each represented by two polynomials) outputs a ciphertext represented by three polynomials, thereby violating the compactness property. This increases the cost of subsequent operations, making it necessary to transform the ciphertext $(d_0, d_1, d_2)$ back into the form $(\bar c_0, \bar c_1)$. To achieve this, a technique known as \textit{relinearization}, or \textit{key switching}, is employed.

It is worth noting that in \cite[Theorem 2]{cryptoeprint:2025/2027}, the authors proposed a way to correctly estimate the coefficients variance of $\nu \cdot \nu'$, which will be further developed and employed later in this work (see \Cref{teo.mul.c2,teo.mul.c1c2}). 

\begin{algobox}{Modulus switching.}
    Given an extended ciphertext $(\bm{c}, q_\ell, \nu)$, and given an integer $q_{\ell'}$ such that $q_{\ell'} < q_\ell$, the modulus switching procedure outputs $(\bm{c}',q_{\ell'},\nums)$, 
where
\begin{equation*}
    \bm{c}' = \frac{q_{\ell'}}{q_\ell} (\bm{c} + \bm{\delta}) \mod q_{\ell'},  \mbox{ with } \bm{\delta} = t[-\bm{c}t^{-1}]_{{q_\ell}/{q_{\ell'}}}.
\end{equation*}
\end{algobox}
\noindent This modulus switching technique allows us to reduce the size of the noise associated with a given ciphertext. Indeed, the output ciphertext still decrypts to the original plaintext (scaled by a factor of ${q_\ell}/q_{\ell'}$), but has a smaller associated noise. For this reason, it is typically applied to ciphertexts involved in homomorphic multiplications, where noise growth is most significant. 

The critical quantity associated with the new ciphertext $\bm{c'}$ can be expressed in terms of that of $\bm{c}$, namely,
\begin{align*}
    \nums &= [c_0'+ c_1' \cdot s]_{q_{\ell'}} =  \frac{q_{\ell'}}{q_\ell}([c_0+c_1 \cdot s]_{q_\ell}+ \delta_0 + \delta_1\cdot s) =  \frac{q_{\ell'}}{q_\ell}(\nu+ \delta_0 + \delta_1\cdot s).
\end{align*}
As shown in \cite{cryptoeprint:2025/2027} the term $\delta_0$ is negligible and the coefficients variance of $\nums $ can be estimated as \begin{align}\label{var.modswitch}
    \V(\nums|_i) &= \left(\frac{q_{\ell'}}{q_\ell}\right)^2\left(\V(\nu|_i) +\V(\delta_1\cdot s|_i) \right)\notag \\
   &= \left(\frac{q_{\ell'}}{q_\ell}\right)^2\left(\V(\nu|_i) + nV_{\delta_1}V_{s}\right).
\end{align}

 \begin{algobox}{Rotations}
     Let $\sigma_k: \R_q \rightarrow \R_q$ be the ring automorphism defined by $\sigma_k(x): x \rightarrow x^k$, where $k$ is an odd integer. Given an extended ciphertext $(\bm{c}, q_\ell, \nu)$ the rotation operation consists in applying the automorphism to both the ciphertext components, i.e., \begin{equation*}
        \sigma_k(\bm{c})= (\sigma_k(c_0),\sigma_k(c_1)) \mod q_\ell.
    \end{equation*} 
\end{algobox}
\noindent The critical quantity associated with the rotated ciphertext is given by \begin{equation*}
    \sigma_k(\nu) = \sigma_k(c_0) + \sigma_k(c_1)\sigma_k(s).
\end{equation*}
Since $\sigma_k(\cdot)$ permutes the coefficients up to signs, the coefficient variance is unchanged, so \begin{equation*}
    \V(\sigma_k(\nu)|_i) = \V(\nu|_j),
\end{equation*}
for a coefficient $j$ established by the automorphism.
Note that the resulting ciphertext encrypts the rotated plaintext with respect to $\sigma_k(s)$. To obtain a ciphertext encrypting the same plaintext but with respect to the initial secret key $s$, the key switching technique is employed. Therefore, throughout this work we will always assume that a key switching operation is performed after rotation. 
This is consistent with the behavior typically observed in the most common FHE libraries. For instance, in OpenFHE, when the operation \texttt{EvalRotate} is used, an implicit key-switching operation is also performed.

Intuitively, the basic idea of key switching is to use an
encryption of a new secret key $\sk'$ under the original secret
key $\sk$ to transform a ciphertext associated with $\sk'$ into
an equivalent ciphertext encrypted under $\sk$.

It is commonly applied in two scenarios. The first occurs after
the multiplication of two ciphertexts: the resulting ciphertext
has three components $(c_0,c_1,c_2)$, and its decryption involves
the secret key powers $1$, $s$, and $s^2$. In this case,
$\sk'=s^2$, and key switching transforms it into an equivalent
ciphertext with only two components,
$\Bar{\bm{c}}=(\bar{c}_0,\bar{c}_1)$, encrypted under the
original secret key $\sk=s$.

The second occurs after a rotation, which transforms the
ciphertext components into $(\sigma_k(c_0),\sigma_k(c_1))$,
resulting in a ciphertext associated with the rotated secret
key $\sk'=\sigma_k(s)$. In this case, key switching transforms
the term $\sigma_k(c_1)\sigma_k(s)$ into an equivalent
representation under the original secret key, where, in this case,  $c_2 = \sigma_k(c_1)$.

\begin{algobox}{Key Switching (Intuitive Idea)}
Let $(\bm{c},q_\ell,\nu)$ be a ciphertext where
$\bm c = (c_0,c_1,c_2)$, and $c_2$ is associated with a secret
key $\sk'$.
The key switching operation generates a key switching key
\begin{equation*}
    \ksk=(\ksk_0,\ksk_1)
    =(-a\cdot\sk+te+\sk',a)\mod q_\ell,
\end{equation*}
where $a\leftarrow\mathcal{U}_{q_\ell}$ and
$e\leftarrow\chi_e$, and outputs
$(\Bar{\bm{c}},q_\ell,\nu_\ks)$, with
\begin{equation*}
    \Bar{\bm{c}}=
    (c_0+c_2\cdot\ksk_0,
    c_1+c_2\cdot\ksk_1)\mod q_\ell.
\end{equation*}
\end{algobox}

Note that, this specific construction is only intended as an informal way of illustrating key switching and is not used in practice, as its error contribution would be excessive. However, in this work, we do not need to specify the details of the method, nor we need to explicitly track the associated error contribution. Indeed, when applied after multiplication, its error contribution can be neglected, see \cite[Appendix C]{cryptoeprint:2025/2027}. For rotations, the key switching operation we consider is always followed by a modulus switching step, which cancels the contribution of the key switching error, as explained in \Cref{sec:theoric_contribution}. Hence, throughout
this work, we abstract away the specific implementation of key switching. 

\section{Average-Case Noise Analysis}\label{sec:theoric_contribution}
In this section, we present our results on how to precisely estimate the variance of the error coefficients arising from the addition and product of \textit{dependent} ciphertexts. 
Most existing average-case analyses rely on the strong assumption that the error terms involved in a computation are independent. However, as shown in 
\cite{cryptoeprint:2025/2027} this does not hold in the BGV context. Indeed, the authors show that even when ciphertexts are independently generated, the associated critical quantities are not independent, since they share structural components induced by the public and secret keys. 
We refer to these dependencies as \textit{common dependencies}, according to the terminology introduced in \cite{cryptoeprint:2025/1036}. Therefore, the authors of \cite{cryptoeprint:2025/2027} provided tight estimation bounds able to capture the effect of these common terms. 
However, beyond these, one must also consider \textit{ciphertext dependencies} \cite{cryptoeprint:2025/1036} which arise when the ciphertexts involved in a computation are themselves dependent. This situation naturally occurs in real-world computations, where operations are often performed on dependent data. 

However, the framework of \cite{cryptoeprint:2025/2027} only discusses common dependencies, and explicitly states that this second source of correlation is not addressed. 
Accounting for both types of dependencies is essential, as they increase the variance of the coefficients of the critical quantities involved. Ignoring them leads to an underestimation of the error, which must be avoided not only to ensure the correct functioning of the scheme but also for security reasons, as shown in \cite{checri2024practical,cheon2024attacks}.
In this section, we fill this gap by providing a precise analysis of the variance of the coefficients of the critical quantity able to account both for common and ciphertext dependencies. 

Furthermore, we explicitly examine how rotations of a ciphertext influence the effects of these dependencies. As discussed in \Cref{sec:poly-eval}, accurately estimating this effect is important not just because rotations can occur within arbitrary computational circuits but also since several recent methods designed to optimize polynomial evaluation directly utilize this operation. Consequently, a precise characterization is essential to the scope of this work.

This section is organized as follows. 
 In \Cref{sec: track} we recall the main results needed to study common dependencies in the case of independent ciphertexts, which will also be used in the analysis of ciphertext dependencies. In \Cref{sec:square}, we derive variance estimates for the critical quantity arising from  products and additions of dependent ciphertexts. Finally, in \Cref{sec: automorphisms}, we accurately estimate the variance resulting from the product of a ciphertext by a rotation of itself. \

\subsection{Tracking Common Dependencies} \label{sec: track}
We briefly recall the main results of \cite{cryptoeprint:2025/2027}, which provide the baseline for our analysis. 
\begin{notation}
    Each critical quantity involved in a circuit of homomorphic operations can be expressed as a polynomial in the secret and public keys, as follows \begin{equation*}
    \nu =  \sum_{\iota} a_{\iota}s^{\iota} = \sum_{\iota}\sum_{\mu} b_{\mu}(\iota) e^{\mu} s^{\iota},
\end{equation*}
where \( a_{\iota} = \sum_{\mu} b_{\mu}(\iota) e^{\mu} \), and \( b_{\mu}(\iota) \) contains no powers of \( s \) or \( e \).
\end{notation}

\begin{lemma}[{\cite[Lemma 3]{cryptoeprint:2025/2027}}]\label{lemma3}
    Let $a(x) = a|_0 +a|_1x+ \dots + a|_{n-1}x^{n-1} \in \mathcal{R}$, where $a|_i$'s are i.i.d. random variables with $\E[a|_i] = 0$,  $\E[a|_i^2] = V_a$. 
Then, as $n$ approaches infinity, for all $0 \leq i \leq n-1$ we have
\begin{equation*}
    \E[a^k|_i]  =0 , \quad \E[(a^k|_i)^2]  = k!n^{k-1}V_a^k \,,
\end{equation*}
 where $a^k|_i$ denotes the $i$--th coefficient of the polynomial~$a(x)^k$.
\end{lemma}
\begin{remark}
    \Cref{lemma3} is an asymptotic statement. However, since BGV typically uses ring dimension $n \ge 2^{13}$, the error introduced by the asymptotic approximation is negligible and the resulting expressions, which we use throughout the paper, can be considered reliable.
\end{remark}

\noindent For clarity of exposition, the main results proved in \cite[Section 6]{cryptoeprint:2025/2027} are summarized in the following statements. 
\begin{lemma}\label{approx_modswitch}
Let $(\bm{c},q_\ell,\nu)$ be an extended ciphertext and let $\nums = \frac{q_\ell'}{q}\bigl(\nu + \delta_1 s\bigr)$,
be the critical quantity after modulus switching to the target modulus $q'_\ell$, where $q = pq_\ell'$.
If $p$ is chosen sufficiently large so that 
$\V(\nu/{p}|_i) \ll \V({\delta_1}s/{p}|_i)$,
then the critical quantity can be approximated as
\begin{equation}\label{err.modswitch}
    \nums \approx \delta_1 s/p,
\end{equation}
where $  \delta_1/p  \leftarrow \mathcal{U}_t$.
In particular, the coefficients of the critical quantity after modulus switching have variance
\begin{equation}\label{Vms}
    \Vms:= nV_{\sfrac{\delta_1}{p}} V_s = n \frac{t^2}{12}V_s\,.
\end{equation}
Thus, the variance after modulus switch no longer depends on the initial ciphertext considered. 
\end{lemma}
In this work, we always assume that whenever modulus switching is applied, the ratio between the moduli is sufficiently large, as required by the condition above. This is a realistic assumption for the parameter choices used in practical applications today, see \cite{cryptoeprint:2025/2027} for an in-depth discussion.

\begin{theorem}\label{independent_variance}
    The product of two independent ciphertexts that have undergone modulus switching has coefficient-wise variance $2n\Vms^2$.
\end{theorem}
Intuitively, this factor 2 corresponds exactly to the correction discussed at the beginning of this section, which accounts for the dependencies introduced by the keys used in the scheme. In particular, we are considering here the product of two ciphertexts whose associated critical quantities are of the form $\delta_1 s/p$ and $\delta'_1 s/p$, where $\delta_1$ and $\delta'_1$ are mutually independent. 
The factor $2$ precisely captures the increase in variance caused by the fact that both expressions share the same secret key $s$. If the two ciphertexts were totally independent, this additional contribution would not appear.
More generally, the correction factor grows with the number of terms shared by the critical quantities involved in the multiplication, as shown in the following section.

\begin{proposition}[{\cite[Proposition 2]{cryptoeprint:2025/2027}}]\label{prop.add}
Let $\bm c$ and $\bm c'$ be two independent ciphertexts with critical quantities $\nu,$ and $\nu'$, respectively. Then, the variance of the error coefficients resulting from the addition of $\bm c$ and $\bm c'$ is 
        \begin{equation}
            \label{eq.V.add}
            \V(\nu_\mathsf{add}|_i) = \V(\nu|_i) + \V(\nu'|_i).
        \end{equation}  
\end{proposition}

\subsection{Tracking Ciphertext Dependencies}\label{sec:square}
In this section, we consider the worst-case scenario in which the number of dependencies is maximized, namely the squaring of a ciphertext.

\begin{theorem}\label{teo.mul.c2}
    Let $(\bm{c},q_\ell,\nums)$ be an extended ciphertext, that has just undergone modulus switching. Let $(\bm{c^{(2)}},q_\ell, \numul)$  be the result of squaring the ciphertext $\bm{c}$. Then, the variance of the coefficients of the resulting critical quantity $\numul$ can be estimated as \begin{equation}
        \V(\numul|_i) \approx 4n\Vms^2
    \end{equation} 
\end{theorem}
\textit{Intuition.} 
In contrast to the independent case, squaring introduces additional dependencies, since both operands share, not only the secret key $s$ as before, but also the same term $\ams$, which is uniformly distributed.  This results in an additional factor $2$ in the variance (see \Cref{independent_variance}).
\begin{proof}
By \Cref{approx_modswitch}, we can write $\nums \approx \ams \cdot s$ and
\begin{equation*}
    \V(\nums|_i)= \Vms = nV_{\ams}V_s.
\end{equation*}
Our aim is to precisely define the variance of the coefficients of the critical quantity $\numul$ associated to $\bm{c^{(2)}}$, i.e. of $\nums^2|_i$. We start by writing $\nums^2|_i$ as 
$$ \nums^2|_i = \left(\ams s\right)^2\Big|_i= \sum_{j=0}^{n-1} \xi(i,j)\left(\ams \right)^2\Big|_j \cdot  s^2|_{i-j}$$ 
We want to estimate 
$$\V\left(\left(\ams s\right)^2\Big|_i\right) = \V\left(\sum_{j=0}^{n-1} \left(\ams \right)^2\Big|_j \cdot s^2|_{i-j}\right) $$
For the sake of clarity, let $X_i =(\ams)^2 |_i$ and $Y_i =s^2 |_i$. We know, according to \Cref{lemma3} that 
    $\E[X_i] = \E[Y_i] = 0\,, \V(X_i) = 2n\Vams^2$, and 
    $\V(Y_i) = 2n\Vs^2$.
We want to study 
\begin{align}\label{sumV}
   \V(\sum_jX_jY_{i-j}) &= \sum_j\V(X_j)\V(Y_{i-j}) + \notag \\ &+\sum_{j_1 \ne j_2}\xi(i,j_1)\xi(i,j_2)\Cov(X_{j_1}Y_{i-j_1},X_{j_2}Y_{i-j_2}) .
\end{align}
It can be proven that $\Cov(X_{j_1}Y_{i-j_1},X_{j_2}Y_{i-j_2})=0$, for $j_1 \ne j_2$. In fact, it is immediate to verify that 
$X_{j_1},  X_{j_2} \perp Y_{i-j_1}Y_{i-j_2}$, and 
$\E[X_{j_1}] = 0$.
Moreover, it is possible to show that $\Cov(X_{j_1},  X_{j_2})=0$ for $j_1 \ne j_2$, thereby concluding $\Cov(X_{j_1}Y_{i-j_1},X_{j_2}Y_{i-j_2})=0$, according to \Cref{probfacts} (1). Indeed,
\begin{align*}
   \Cov(X_{j_1},  X_{j_2}) &= \E[X_{j_1}X_{j_2}] = \E\left[\left(\ams\right)^2 \Big|_{j_1}\left(\ams\right)^2 \Big|_{j_2}\right]= \\ &= \frac{1}{p^4}  \E[\sum_{k_1}\xi(j_1,k_1) \delta_1 |_{k_1}\delta_1 |_{j_1-k_1}\sum_{k_2} \xi(j_2,k_2) \delta_1 |_{k_2}\delta_1 |_{j_2-k_2}] = \\
   &=  \frac{1}{p^4}\sum_{k_1}\sum_{k_2}\xi(j_1,k_1) \xi(j_2,k_2)\E[ \delta_1 |_{k_1} \delta_1 |_{j_1-k_1} \delta_1 |_{k_2}\delta_1 |_{j_2-k_2}] .
\end{align*} 
We are considering $j_1 \ne j_2$, hence we have two possible cases: $k_1 = k_2$ or $k_1 \ne k_2$.
\begin{itemize}
    \item When $k_1 = k_2 = k$ then $j_1-k \ne j_2-k$, hence at least one among  $\delta_1 |_{j_1-k},\delta_1 |_{j_2-k}$ is different (and so independent) from $\delta_1 |_{k}$, thus  
\begin{equation*}
    \E[ \delta_1 |_{k}^2\delta_1 |_{j_1-k}\delta_1 |_{j_2-k}] = \E[ \delta_1 |_{j_1-k}]\E[  \delta_1 |_{k}^2 \delta_1 |_{j_2-k}] = 0.
\end{equation*}
\item
When $k_1 \ne k_2 $ then $\xi(j_1,k_1) \xi(j_2,k_2) \E[ \delta_1 |_{k_1} \delta_1 |_{j_1-k_1} \delta_1 |_{k_2} \delta_1 |_{j_2-k_2}]$.
Again, if one term is different from the others, the total value is zero. Therefore, the only way this can be different from zero is if the indexes are pairwise equal (modulo n). In such case we get $\E[ \delta_1 |_{k_1}^2 \delta_1 |_{k_2}^2]$. Again, we have two possibilities \begin{itemize}
    \item $k_1 = j_2-k_2$ and $k_2 = j_1-k_1$, which however means that $j_1 = j_2$ which is a contradiction and therefore cannot occur.
    \item $k_1 = j_1-k_1$ and $k_2 = j_2-k_2$, which means that $j_1 = 2k_1 \mod n$ and $j_2 = 2k_2 \mod n$. Since $n$ is even, if there is a solution $k_i ={j_i}/{2}$ then also  $k_i ={j_i}/{2} +  n/2$
    is a solution. For one of them, we obtain a positive sign $\xi(j_1,k_1)=1$, whereas for the other we obtain a negative sign $\xi(j_1,k_1)=-1$. As a consequence, the two contributions cancel each other out, yielding a total contribution of zero also in these cases.
\end{itemize}
\end{itemize}
Therefore, according to \Cref{probfacts} (2), \Cref{sumV} becomes 
\begin{align*}
   \V(\sum_jX_jY_{i-j}) &= \sum_j\V(X_jY_{i-j}) = \sum_j\V(X_j)\V(Y_{i-j}) \\& = n(2n\Vams^2\cdot 2n\Vs^2)= 4n(n\Vams\cdot\Vs)^2 = 4n\Vms^2,
\end{align*}
where $\Vms$ is as in \Cref{Vms}.
\qed
\end{proof}

When considering higher powers of the same ciphertext, one would naturally expect a significant increase in dependencies among the resulting terms. Indeed, repeated multiplications of the same ciphertext introduce a large number of shared monomials in the underlying polynomial representation of the critical quantity. These induce additional correlations which, in principle, should lead to a more intricate variance structure and potentially larger coefficient growth,  as happens, for example, in the BFV scheme \cite{BFVvar}. 

However, the effect of these dependencies is substantially simplified in real-world scenarios, thanks to the use of modulus switching. As shown in \Cref{approx_modswitch}, after modulus switching the critical quantity admits the approximation
    \begin{equation*}
    \nums \approx \delta_1 s/p.
\end{equation*}
where $\ams$ is uniformly distributed and, crucially, is no longer affected by dependencies caused by the secret/public keys or, most importantly for our analysis, by the randomness shared among dependent ciphertexts. 

\begin{theorem}\label{teo.mul.c1c2} 
    Let $(\bm{c}_1,q,\nu_1)$ and $(\bm{c}_{2},q,\nu_2)$ be two (possibly dependent) extended ciphertexts that have just undergone modulus switching. Let $(\bm{c}_{\mathsf{mul}},q,\nu_{\mathsf{mul}})$ denote the resulting ciphertext. 
    Then, \begin{equation*}
        \V(\nu_{\mathsf{mul}}|_i) \le 4n\Vms^2.
    \end{equation*}
\end{theorem}

\begin{proof}
Since both $\bm{c}_1$ and $\bm{c}_2$ have just undergone modulus switch, according to \Cref{approx_modswitch} the associated critical quantities are of the form
    \begin{align*}
       \nu_1 \approx  \frac{1}{p_1}{\delta_1}^{(\bm{c_1})} \cdot  s \quad \mbox{ and }\quad 
       {\nu_2} \approx  \frac{1}{p_2}{\delta_1}^{({\bm{c_2}})} \cdot  s .
    \end{align*}
   Define $X_1 =\frac{{\delta_1}^{(\bm{c_1})}}{p_1} , X_2=  \frac{{\delta_1}^{({\bm{c_2}})}}{p_2}, Z=X_1\cdot X_2, S= s^2$, it holds that, \begin{equation*}
        \nu_{\mathsf{mul}} = Z \cdot S,
    \end{equation*}
    and we are interested in bounding the following quantity \begin{equation*}
        \V(Z\cdot S|_i)= \V(\sum_{j=0}^{n-1} \xi(i,j) Z|_j\cdot S|_{i-j}) =\sum_{j=0}^{n-1}\V(Z|_j\cdot S|_{i-j}) ,
    \end{equation*} where the last equality holds since all the covariance terms vanish by \Cref{probfacts}~(1). Indeed, $S$ is independent of $Z$, $\E[S|_{i-j_1}]=0$ by
\Cref{lemma3}, and $\Cov(S|_{i-j_1},S|_{i-j_2})=0$, as shown in the proof of
\Cref{teo.mul.c2}.
    Using again that $S$ and $Z$ are independent, and that $\E[S|_i^2]=2nV_s^2, \E[S|_i]=0$ by \Cref{lemma3} we obtain \begin{equation*}
     \V(Z|_j\cdot S|_{i-j}) = \E[Z|_j^2]\cdot \E[S|_{i-j}^2] -(\E[Z|_i]\cdot \E[S|_{i-j}])^2 = 2nV_s^2\cdot \E[Z|_j^2],
    \end{equation*}
    thus, \begin{equation*}
       \V(\nu_{\mathsf{mul}}|_i)=\V(Z\cdot S|_i)=\sum_{j=0}^{n-1}2nV_s^2\cdot \E[Z|_j^2] = 2nV_s^2  \sum_{j=0}^{n-1}\E[Z|_j^2]  .
    \end{equation*}
    It remains to show that $\sum_{j=0}^{n-1}\E[Z|_j^2]  \le n\left(2nV_{X_1}^2\right)$. In such case,   \begin{equation*}
       \V(\nu_{\mathsf{mul}}|_i)\le  4n (n^2V_s^2 V_{X_1}^2) =4n\Vms^2  .
    \end{equation*}
    Using \Cref{lem:parseval}, \Cref{probfacts} (5) and that the canonical embedding is a homomorphism, it holds that,  \begin{align*}
        \sum_{j=0}^{n-1}\E[Z|_j^2] &= \frac{1}{n} \sum_{m=1}^n\E[|Z(\zeta_m)|^2]= \frac{1}{n} \sum_{m=1}^n\E[|X_1(\zeta_m)|^2 \cdot |X_2(\zeta_m)|^2]\\
        &\le \frac{1}{n} \sum_{m=1}^n\sqrt{\E[|X_1(\zeta_m)|^4 ]\cdot \E[|X_2(\zeta_m)|^4]}= \frac{1}{n} \sum_{m=1}^n\E[|X_1(\zeta_m)|^4 ]\\
         &= \frac{1}{n} \sum_{m=1}^n\E[|X_1^2(\zeta_m)|^2 ]= \sum_{j=0}^{n-1} \E[X_1^2|_j^2 ] = n(2nV_{X_1}^2),
    \end{align*}
    thus concluding the proof.
\end{proof}

\noindent With the purpose of evaluating any function of the form $f(x)=\sum_{j=1}^\kappa a_j x^j$, we also need to introduce the result below, regarding the addition of (possibly) dependent ciphertexts.

\begin{proposition}\label{prop.add.dep}
Let $(\bm c_1, q_1, \nu_1)$ and $(\bm c_2, q_2, \nu_2)$ be two (possibly dependent) extended ciphertexts. Then the variance of the error coefficients $\nu_{\mathsf{add}}$ resulting from the addition of $\bm c_1$ and $\bm c_2$ is 
    $$
    \V(\nu_{\mathsf{add}}|_i)\le \V(\nu_1|_i) + \V(\nu_2|_i) + 2 \sqrt{\V(\nu_1|_i) \V(\nu_2|_i)}
    $$
In particular, let $\kappa_1, \kappa_2$ be two distinct positive integers. If $\bm{c}_1 = \bm c^{\kappa_1}$ and $\bm{c}_2 = \bm c^{\kappa_2}$ are obtained by computing powers of a ciphertext $\bm{c}$, then \[\V(\nu_{\mathsf{add}}|_i) \approx \begin{cases} \V(\nu_1|_i) + \Vms  \quad \text{if $q_1 < q_2$} \cr \V(\nu_1|_i) + \V(\nu_2|_i) \quad \text{if $q_1 = q_2$}  \end{cases}.\]
\end{proposition}

\begin{proof}
From point (3) of \Cref{probfacts}, it holds that \begin{equation*}
   \V(\nu_{\mathsf{add}}|_i) = \V(\nu_1|_i) + \V(\nu_2|_i) + 2\Cov(\nu_1|_i,\nu_2|_i).
\end{equation*}   
Since $\V(\nu_1|_i)$ and $\V(\nu_2|_i)$ are finite, then 
$\Cov(\nu_1|_i, \nu_2|_i)\leq \sqrt{\V(\nu_1|_i) \V(\nu_2|_i)}$.
In particular, let  $\bm{c}_1 = \bm c^{\kappa_1}$ and $\bm{c}_2 = \bm c^{\kappa_2}$ with $\kappa_1\ne\kappa_2$.
\begin{itemize}
    \item If $q_1 < q_2$, then $\bm c^{\kappa_2}$ is first subjected to a modulus switch to $q_1$. Hence, according to \Cref{approx_modswitch} the variance of the critical quantity can be approximated by $\Vms$. Note that we always assume to perform a modulus switch before multiplication, thus $\nu_1$ is the product of terms as in \Cref{err.modswitch}. By applying the same argument as in \Cref{teo.mul.c2}, it is straightforward to see that the covariance is null, which concludes the first part of the proof.
     \item If $q_1 = q_2$, then there is no need to apply modulus switching to $q_1$. Now, both  $\nu_1, \nu_2$ are products of terms as in \Cref{err.modswitch} and again by applying the same reasoning it can be shown that the covariance vanishes, concluding the proof.
\end{itemize} \qed
\end{proof}

\subsection{Rotations}\label{sec: automorphisms}
Finally, we study how dependencies influence the variance resulting from the product of a ciphertext with one of its rotations. This scenario is of particular interest not just because it may appear in a generic homomorphic evaluation circuit, but especially because some recent methods that optimize polynomial evaluation \cite{10.1007/978-981-99-8736-8_3,cryptoeprint:2026/1089} directly involve computations of this form, as will be discussed later. In particular, one may ask whether the product of a ciphertext and its rotation behaves more like the product of two independently generated ciphertexts or like the squaring of a single ciphertext. In this section, we show that modulus switching, combined with key switching, makes the product of a ciphertext and its rotation resemble the first scenario. To introduce this result, formalized in \Cref{thm:autoMS}, we first introduce an auxiliary lemma. 

\begin{lemma}\label{lemma:surjective}
    Let $p$ be a prime and let $J\subseteq\R_p$ be an ideal with
$\dim_{\mathbb{F}_p}J>n/2$. Then, for every pair of distinct indices
$j_1\neq j_2$, the coordinate projection
\[
   \pi:J\longrightarrow\mathbb{F}_p^2,\qquad a\longmapsto\bigl(a|_{j_1},a|_{j_2}\bigr)
\]
is surjective, i.e. $\pi(J)=\mathbb{F}_p^2$.
\end{lemma}
\begin{proof}
    We refer the reader to Appendix \ref{appendix-lemma:surjective} for the proof of the lemma.
\end{proof}

\begin{theorem}\label{thm:autoMS}
Let $k$ be an odd integer and let $\sigma_k: \R_{pq}\mapsto \R_{pq}$ be the ring automorphism defined by $\sigma_k: x \mapsto x^k$. Let $(\mathbf c,pq,\nu)$ be an
extended ciphertext. 
Assume that $\sigma_k(\mathbf c)$ first undergoes key switching to the original secret key, yielding $\Bar{\bm c}$. Then $\mathbf c$ and $\Bar{\bm c}$ undergo modulus switching and are multiplied. Denote by $(\mathbf c_{\sigma_k},q,\nu_{\sigma_k})$ the result of these operations. Then
\[
\mathrm{Var}\big(\nu_{\sigma_k}|_i\big)\approx 2n\,V_{ms}^2 .
\]
\end{theorem}

\begin{proof}
    We start by observing that before modulus switching, the rotated ciphertext must go through a key switching procedure, since $\sigma_k(\mathbf c)$ is a ciphertext under the rotated secret key $\sigma_k(s)$. However, we do not need to explicitly compute the noise introduced by key
    switching. Indeed,  by \Cref{approx_modswitch} we know that, after modulus switch, the dominant component of the error is given by the correction term introduced by modulus switching itself (i.e. the term of the form $\delta s$), which dominates the contribution of the previous noise, making it negligible.  
    Thus, we are going to consider the very naive key switching technique, which simply consists in encrypting the old secret key. 
    After key switching, the rotated ciphertext is given by
    \begin{equation*}
         \Bar{\bm c}=(\sigma_k(c_0)+\sigma_k(c_1)\cdot \ksk_0,\sigma_k(c_1)\cdot \ksk_1) \mod pq,
    \end{equation*}
    where \begin{equation*}
           \ksk=(\ksk_0,\ksk_1)= (-a_\ks  s +te_\ks + \sigma_k(s),a_\ks),
    \end{equation*}
    is the key switching key and  $a_\ks \leftarrow \mathcal{U}_{pq}$ and $e_\ks \leftarrow \mathcal{\chi}_e$.
Before multiplication, both $\bm c$ and $\Bar{\bm c}$ are modulus switched to $q$.
     Denote by  
       $$
       \bm{\delta}^{(\bm{c})}:= t[-\bm c t^{-1}]_{p} \quad \mbox{ and }\quad 
       \bm{\delta}^{(\Bar{\bm c})}:= t[-\Bar{\bm c}t^{-1}]_{p},
       $$
    the correction terms introduced by modulus switching, where, by construction, the quantities $[-\bm c t^{-1}]_{p}$ and $[-\Bar{\bm c}t^{-1}]_{p}$ are RLWE samples. 
    By \Cref{approx_modswitch}, after modulus switching to $q$, the critical quantities associated with the two ciphertexts are given by
    \begin{align*}
       \nu \approx  \frac{1}{p}{\delta_1}^{(\bm{c})} \cdot  s \quad \mbox{ and }\quad 
       \Bar{\nu} \approx  \frac{1}{p}{\delta_1}^{(\Bar{\bm{c}})} \cdot  s .
    \end{align*}
Denote by $X_i \coloneqq \frac{1}{p^2} {\delta_1}^{(\Bar{\bm{c}})}{\delta_1}^{({\bm{c}})}|_i $, $Y_i \coloneqq s^2|_i $. Following the same reasoning of \Cref{teo.mul.c2}, the variance of the coefficients of $\nu_{\sigma_k}=\nu \cdot \Bar{\nu}$ can be written as \begin{align*}
   \V(\nu_{\sigma_k}|_i) &= \V(\sum_jX_jY_{i-j}) = \sum_j\V(X_j)\V(Y_{i-j}) = n(\V(X_j)\cdot 2n\Vs^2).
\end{align*} 
    Hence, the proof consists in showing that \begin{equation*}
        \V(X_j)= \V( \frac{1}{p}{\delta_1}^{(\Bar{\bm{c}})} \cdot\frac{1}{p}{\delta_1}^{({\bm{c}})} |_j) = nV_{\frac{1}{p}\delta_1^{(\Bar{\bm{c}})}}V_{\frac{1}{p}\delta_1^{({\bm{c}})}},
    \end{equation*}
    which is equivalent to prove that \begin{equation*}
        \sum_{j_1 \neq j_2}\Cov({\delta_1}^{(\Bar{\bm{c}})}|_{j_1}{\delta_1}^{({\bm{c}})} |_{i-j_1},{\delta_1}^{(\Bar{\bm{c}})}|_{j_2}{\delta_1}^{({\bm{c}})} |_{i-j_2})  = 0,
    \end{equation*}
    i.e., \begin{equation*} \E[{\delta_1}^{(\Bar{\bm{c}})}|_{j_1}{\delta_1}^{({\bm{c}})} |_{i-j_1}{\delta_1}^{(\Bar{\bm{c}})}|_{j_2}{\delta_1}^{({\bm{c}})} |_{i-j_2}] = 0 \quad \text{for } j_1 \neq j_2.
    \end{equation*}
    Using the tower property, we can write  \begin{equation*} \E[{\delta_1}^{(\Bar{\bm{c}})}|_{j_1}{\delta_1}^{({\bm{c}})} |_{i-j_1}{\delta_1}^{(\Bar{\bm{c}})}|_{j_2}{\delta_1}^{({\bm{c}})} |_{i-j_2}] = \E\!\left[{\delta_1}^{({\bm{c}})} |_{i-j_1}{\delta_1}^{({\bm{c}})} |_{i-j_2}\E[{\delta_1}^{(\Bar{\bm{c}})}|_{j_2}{\delta_1}^{(\Bar{\bm{c}})}|_{j_1} \,\big| \, \bm{c}]\,\right]\!.  \end{equation*}
Focusing on $${\delta_1}^{(\Bar{\bm{c}})} =t[- t^{-1}\sigma_k(c_1)a_\ks]_{p}, $$ we first observe that, conditionally on $\bm c$ the element
    $\sigma_k(c_1)$ is fixed, while $a_\ks \leftarrow \U_{pq}$ is freshly generated during the key switching procedure. 
For a fixed $\sigma_k(c_1)$, multiplication by $\sigma_k(c_1)$ is an 
    $\R_p$-linear map
    \begin{equation*}
          m_{\sigma_k(c_1)}:\R_p\longrightarrow \R_p,\qquad x\longmapsto \sigma_k(c_1)x.
    \end{equation*}
    Hence, since $a_\ks \mod p\leftarrow \mathcal U_p$,  $\sigma_k(c_1)\cdot a_\ks$ is uniform over the image of $m_{\sigma_k(c_1)}$, that coincides with the  principal ideal $J \coloneqq \sigma_k(c_1)\R_p$.  Fix $j_1\neq j_2$ and let $\pi:\R_p\to\mathbb{F}_p^2$ be the projection onto the coordinates $j_1,j_2$. The composition $\pi\circ m_{\sigma_k(c_1)}$ is
    $\mathbb{F}_p$-linear, hence a homomorphism of finite abelian groups. Using \Cref{lemma:surjective} we know that for every parameter
    set considered in this work $\pi(J)=\mathbb{F}_p^2$ with overwhelming probability. Indeed, $\dim_{\mathbb{F}_p}J = n -\deg g$ where $g \coloneqq \gcd(\sigma_k(c_1),x^n+1)$. Hence, \begin{equation*}
        \dim_{\mathbb{F}_p}J > \frac{n}{2} \Leftrightarrow \deg g < \frac{n}{2}
    \end{equation*}
    However, \begin{equation*}
        \mathbb{P}[\deg g \ge \frac{n}{2}]  \le \!\!\!\!\sum_{\substack{
g \mid x^n+1\\
\ \deg g \ge n/2
}}\!\!\!\!\mathbb{P}[g \mid \sigma_k(c_1)] =\!\!\!\! \sum_{\substack{
g \mid x^n+1\\
\ \deg g \ge n/2
}}\!\!\!\! p^{-\deg g} \le 2^np^{-n/2} =   2^{-n/2(\log_2p-2)},
    \end{equation*}
    which is negligible with the common choice of parameters. 
    This means that we can apply \Cref{lemma:surjective} and assume the two coefficients ${\delta_1}^{(\Bar{\bm{c}})}|_{j_2},{\delta_1}^{(\Bar{\bm{c}})}|_{j_1}$ to be independent and uniform over $\mathbb{F}_p$,    i.e., \begin{equation*}
        \E[{\delta_1}^{(\Bar{\bm{c}})}|_{j_2}{\delta_1}^{(\Bar{\bm{c}})}|_{j_1} \,\big| \, \bm{c}] = \E[{\delta_1}^{(\Bar{\bm{c}})}|_{j_2}{\delta_1}^{(\Bar{\bm{c}})}|_{j_1}] = \E[{\delta_1}^{(\Bar{\bm{c}})}|_{j_2}]\E[{\delta_1}^{(\Bar{\bm{c}})}|_{j_1}] =0,
    \end{equation*}
 thus concluding the proof. 
\end{proof}

\begin{remark} \label{rem:epsi}
 The results presented so far are based on the assumption that the ciphertext moduli ratio is large enough to ensure that \Cref{approx_modswitch} holds. However, for parameter selection, it is necessary to quantify how large this ratio actually needs to be. Let $\nu$ be the initial critical quantity and  $\nums = \frac{q_\ell'}{q}\bigl(\nu + \delta_1 s\bigr)$ the corresponding quantity after modulus switching. According to \cite{cryptoeprint:2025/2027}, we define a parameter $\alpha$ such that $\V(\nu/{p}|_i) < \alpha \V({\delta_1}s/{p}|_i)$, where we set $\alpha = \sfrac{1}{100}$. In this way, we ensure that the error introduced by modulus switch dominates the one derived from the previous error term, while still imposing conditions on the ratio of the moduli that are not too restrictive and allow us to provide an improvement in the choice of the parameters.

   As emphasized throughout this work, avoiding underestimation is essential to ensure the correct functioning of the scheme and preserve its security. Since our estimates are very close to the experimental values, we suggest to incorporate the margin factor $\epsilon = 6\alpha$ in the variance estimates of \Cref{teo.mul.c2,teo.mul.c1c2,thm:autoMS}, to account for the residual contribution of the pre-modulus-switching noise.  
   Hence, the variance becomes $(4+\epsilon)n\Vms^2$ (\Cref{teo.mul.c2,teo.mul.c1c2}) and $(2+\epsilon)n\Vms^2$  (\Cref{thm:autoMS}). The rationale behind this correction is explained in depth in \cite{cryptoeprint:2025/2027}. However, the idea is the following. After modulus switching, ciphertext multiplication involves products of critical quantities of the form $\nums = \frac{q'}{q}\bigl(\nu + \delta s\bigr)$. In the worst-case scenario, where the two quantities are equal (maximum dependency) one must also account for the cross terms between $\nu$ and $\delta s$. Moreover, since modulus switching is applied before each multiplication step, the highest power of $s$ appearing in $\nu$ can be assumed to be at most quadratic, hence it is sufficient to use as correction factor $F(2,1)$ \cite[Definition 4]{cryptoeprint:2025/2027}, which leads precisely to the given choice of $\epsilon$. We observe that, for $\alpha = \sfrac{1}{100}$, the induced correction is very small.\ec
\end{remark}

\section{Variance Estimates for Polynomial Evaluation}\label{sec:poly-eval}
\noindent  The results presented in \Cref{sec:theoric_contribution} provide the building blocks for estimating the noise growth in the evaluation of an arbitrary polynomial over a ciphertext. The resulting variance depends on the evaluation strategy, since different strategies may produce different multiplicative depths and apply modulus switching at different stages.

A key insight of our work is that the variance of the coefficients after polynomial evaluation is ultimately determined by (i)  the multiplicative depth associated with the computation of each power; (ii) the terms that, in the final addition, do not undergo a final modulus switch (see \Cref{prop.add.dep}).
These two factors determine which variance estimates from \Cref{sec:theoric_contribution} contribute to the final result.

This observation provides a general methodology for analyzing polynomial evaluation strategies: once the sequence of homomorphic operations and modulus-switching steps is specified, the variance can be estimated by tracking the contributions of the individual terms and their final addition. In the following, we illustrate this methodology through two standard strategies, square-and-multiply and baby-step giant-step, and discuss its applicability to rotation-based methods, which additionally involve products of a ciphertext with its rotations.

\subsection{Square and Multiply}\label{sec.SM}
The following results provide a concrete instantiation of the method presented so far,
giving an accurate estimate of the noise derived from the evaluation of a polynomial on a
ciphertext, using the square-and-multiply algorithm. We chose to compute each power of a ciphertext $\bm c$ using the \emph{binary method} \cite{Knuth2}, which means that we apply recursively \begin{equation*}
    \bm c^j=\bm c^{2^{\lfloor\log_2 j\rfloor}}\bm c^{j-2^{\lfloor\log_2 j\rfloor}}.
\end{equation*} 
In this way, monomials whose exponents are powers of 2 are obtained by multiplying a ciphertext by itself, while other powers are given by the product of two distinct powers. Estimates of the critical quantities associated with these computations are covered by \Cref{teo.mul.c2,teo.mul.c1c2}, respectively. Note that the result in \Cref{teo.mul.ckpow.sm} holds unchanged when powers are computed using \emph{balanced} binary multiplication trees. 
\begin{theorem}\label{teo.mul.ckpow.sm}
Let $(\bm{c},q,\nu)$ be an extended ciphertext and let $f(x)=\sum_{j=1}^\kappa a_j x^j$ be a
polynomial, with $\kappa\geq 2$ and $d = \lceil \log_2 \kappa \rceil$. Let $\bm{c}_f$ denote
the ciphertext obtained by evaluating $f$ on $\bm c$ and let $\nu_f$ be the associated critical quantity. Then, using the square-and-multiply
algorithm, we have
\begin{align*}
    \V(\nu_{f}|_i) \leq &\, m\Vms +
      \sum_{\lceil\log_2 j\rceil = d} 4n a_j^2 \Vms^2. 
\end{align*}
where $m$ is the number of monomials of $f(x)$ of degree at most $2^{d-1}$, namely,
$m=\#\{1\le j\le2^{d-1}:\ a_j\ne0\}$.
\end{theorem}

\begin{proof}
For a fixed power $j$, consider the corresponding monomial of $f$ evaluated in $\bm c$, namely $a_j\bm c^j$. Then,
\begin{itemize}
    \item if $j$ is a power of $2$, then $\bm c^j= \bm c^{j/2}\bm c^{j/2}$, thus the coefficient variance of the associated critical quantity  is approximately $4n\Vms^2$ by
    \Cref{teo.mul.c2}.
    \item if $j$ is not a power of $2$, then $\bm c^j= \bm c^{k}\bm c^{l}$, with $k \neq l$, hence
    the two factors are distinct powers of $\bm c$. 
    Thus,  the coefficient variance of the associated critical quantity is upper bounded by $4 n\Vms^2$ according to \Cref{teo.mul.c1c2}.
\end{itemize}

\noindent At this point, to perform the addition, the ciphertexts must lie in the same ring. Hence,

\begin{itemize}
\item if $\lceil\log_2 j\rceil<d=\lceil\log_2\kappa\rceil$, i.e.
    $j\le2^{d-1}$, modulus switching is applied to
    $a_j\bm c^{\,j}$, yielding an associated critical quantity equal to $\nums$ with coefficient variance $\Vms $ (\Cref{approx_modswitch}).

    \item if $\lceil \log_2 j \rceil= d$, modulus switching is not applied to the ciphertext before the addition, hence the contribution of the ciphertext to the final variance is upper bounded by $4n a_j^2 \Vms^2$, using the fact that $V_{\gamma c} = \gamma^2 V_c$,  $\forall \gamma\in\Z, c\in\R$.
\end{itemize}
The claim follows straightforwardly by summing all ciphertexts using \Cref{prop.add.dep} and observing that the number of ciphertexts that are modulus switched before addition is exactly $m$. 
\qed
\end{proof}

\noindent  A simplified estimate is provided in the following corollary.

\begin{corollary}\label{teo.mul.ckpow} 
Let $(\bm{c},q,\nu)$ be an extended ciphertext and let  $f(x)=\sum_{j=1}^\kappa a_j x^j$ be a polynomial, with $\kappa\geq 2$ and $d = \lceil \log_2 \kappa \rceil$. Let $\bm c_f$ denote the ciphertext obtained by evaluating $f$ on $\bm c$ and let $\nu_f$ be the associated critical quantity.
Then,
\begin{align*}
    \V(\nu_{f}|_i) \leq &\, 2^{d-1}\Vms + \sum_{\lceil\log_2 j\rceil = d} 4n a_j^2 \Vms^2  \approx 4n\Vms^2\sum_{\lceil\log_2 j\rceil = d} a_j^2. 
\end{align*}
\end{corollary}

\begin{proof}
The claim follows immediately by \Cref{teo.mul.ckpow.sm}, by observing that: $(i)$ $m$ is by definition at most $2^{d-1}$, $(ii)$ in the concrete usage of BGV, $n \ge 2^{12}$, hence, the first term is negligible with respect to the second.   \qed
\end{proof}

\noindent In other words, \Cref{teo.mul.ckpow} shows that only the monomials not refreshed by modulus switching contribute to the final variance estimate, namely those requiring the maximum multiplicative depth among the terms of $f$. For example, when evaluating $x^8+x^4$, only the contribution of $x^8$ appears in the estimate, since the ciphertext associated with $x^4$ must undergo an additional modulus-switching before being added to $x^8$. In contrast, for $x^8+x^5$, both terms contribute to the final variance.

\subsection{Baby-Step Giant-Step}
The next statement gives an instantiation of our method when the baby-step giant-step algorithm \cite{paterson1973number} is employed. Let $f(x)=\sum_{j=1}^{\kappa}a_jx^j$ be a polynomial of degree
$\kappa\geq 2$, and set $m=\lceil\sqrt{\kappa+1}\,\rceil$.
The polynomial can be decomposed as
\begin{equation}\label{eq:bsgs}
  f(x)=\sum_{j=0}^{\lfloor\kappa/m\rfloor}
  (x^m)^j B_j(x),
  \qquad
  B_j(x)=\sum_{i=0}^{m-1}a_{jm+i}x^i,
\end{equation}
following~\cite[Section 4]{10.1007/978-981-99-8736-8_3}.
In our analysis, each giant power $(\bm c^m)^j$ is computed using a balanced binary multiplication tree, rather than combining the blocks as in~\cite{wang2025novel}.  
It is worth noting that, even for a fixed polynomial algorithm, the noise estimate also depends on implementation choices such as the order of the homomorphic operations and when modulus switching is applied throughout the computation.

\begin{theorem}\label{teo.bsgs}
Let $(\bm c,q,\nu)$ be an extended ciphertext and let
$f(x)=\sum_{j=1}^{\kappa}a_jx^j$ be a polynomial of degree $\kappa\geq 2$. Set
\[
m=\lceil\sqrt{\kappa+1}\,\rceil,
\qquad
\kappa=\rho m+\beta,
\qquad 0\leq\beta<m.
\]
Let $\bm c_f$ be the ciphertext obtained by evaluating $f$ on $\bm c$ using the baby-step giant-step algorithm described above, and let $\nu_f$ denote its associated critical quantity. Finally, define
\[
A=
\begin{cases}
2^{\lceil\log_2\rho\rceil-1}, & \text{if }\rho\geq 2,\\
0, & \text{if }\rho=1.
\end{cases}
\]
If $\beta\neq 0$, then
    \[
    \V(\nu_f|_i)
    \leq
    (\rho+1)\Vms+
    4n(\rho-A)\Vms^2 \,.
    \]

\noindent If $\beta=0$, then
\[ 
\V(\nu_f|_i)\le
\begin{dcases} \\[-3ex]
\Vms + 4n\,a_\kappa^2\,\Vms^2,
  & \rho = 1,\\
2\Vms + 4n\bigl(1+a_\kappa^2\bigr)\Vms^2,
  & \rho = 2,\\
(\rho+1)\Vms + 4n(\rho-A-1)\Vms^2,
  & \rho\ge 3 \text{ and } (\star),\\
(A+1)\Vms + 2n\Bigl(A+2\sum_{A<j\le\rho} a_{jm}^2\Bigr)\Vms^2,
  & \text{otherwise},
\end{dcases}
\]
where $(\star)$ indicates that there exists $j$ with $(A+1)m<j<\kappa$,
$m\nmid j$ and $a_j\neq 0$.
\end{theorem}

\begin{sketchofproof}
    First, we write the polynomial as in \Cref{eq:bsgs}, that is $f(x)=\sum_{j=0}^{\rho}x^{jm}B_j(x)$ and we evaluate each block separately. For a block with $j \ge 2$, the degrees appearing in $B_j$ are smaller than $m$, meaning that the corresponding baby-step powers must undergo modulus switch before being multiplied by the giant power $\bm c^{jm}$. Hence, each nonzero term in the block has a variance contribution bounded by $\Vms$, while the multiplication by the giant power contributes at most $4n\Vms^2$.
Finally, when all blocks are added, those below the maximum multiplicative level undergo an additional modulus switching, and contribute at most $\Vms$ each, while blocks at maximum level retain the
contribution of their products. Depending on their structure, this contribution is bounded by $4n\Vms^2$ or, when a block reduces to a single monomial $a_{jm}x^{jm}$, by $4n a_{jm}^2\Vms^2$ (which can only occur if $\beta =0$).
The blocks $B_0$ and $B_1$ require separate consideration, since they do not follow the same pattern as the remaining blocks. In particular, the resulting bound depends on whether
$\beta=0$ and on which coefficients of $f$ are nonzero.
Counting the contributions at the maximal level concludes the proof. The complete proof is provided in Appendix \ref{proof.teo.bsgs}. \qed
\end{sketchofproof} 

\begin{corollary}\label{cor.bsgs}
Let $(\bm c,q,\nu)$ be an extended ciphertext, and let $f(x)=\sum_{j=1}^{\kappa}a_jx^j$ be a 
polynomial of degree $\kappa\ge2$, where $\kappa=\rho m+\beta$,  
$m=\lceil\sqrt{\kappa+1}\,\rceil$ and $0\leq \beta<m$. Let $\bm c_f$ be the ciphertext obtained by evaluating $f$ on $\bm c$ using the baby-step giant-step algorithm, and let $\nu_f$ denote its critical quantity. Then,
\[
  \V(\nu_{f}|_i)\le(\rho+1)\Vms
  +4n\Bigl(\Bigl\lceil\frac\rho 2\Bigr\rceil+\delta\sum_{A<j\le\rho}a_{jm}^{2}\Bigr)\Vms^{2},
\]
where $\delta=1$ if $\beta=0$ and $0$ otherwise.
\end{corollary}

\begin{proof}
If $\beta\neq0$, the bound in
\Cref{teo.bsgs} can be directly upper bounded using that $\rho-A\leq\lceil\rho/2\rceil$.
If $\beta=0$, the first two cases of \Cref{teo.bsgs} are covered by the claimed bound. The case $\rho\geq3$ holds since
$\rho-A-1\leq\lceil\rho/2\rceil$, while for the remaining case, we can use that
\[
A+2\sum_{A<j\leq\rho}a_{jm}^2
\leq
2\left(\left\lceil\frac{\rho}{2}\right\rceil+
\sum_{A<j\leq\rho}a_{jm}^2\right).
\] \qed
\end{proof}
\Cref{cor.bsgs} provides a simpler and more compact bound, which is close to \Cref{teo.bsgs} when $\beta\ne0$ or 
when the coefficients $a_{jm}$, $A<j\le\rho$, are small. However, if $\beta=0$ and some of these coefficients are large, it may considerably overestimate the variance: for instance, for 
$f(x)=128x^{72}+x^{64}$ the gap is about $12$ bits, since in this case the coefficient $a_{72}$ is 
absorbed by modulus switching. In such cases \Cref{teo.bsgs} should be preferred.

\paragraph{Choosing the algorithm.}
A further consequence of our framework is that it provides a systematic way to compare different polynomial-evaluation strategies in BGV. Indeed, once the sequence of homomorphic operations and modulus-switching steps is fixed, our results allow one to estimate the corresponding noise growth and, consequently, to choose the evaluation strategy that optimizes performances. For square-and-multiply (SM) and baby-step giant-step (BSGS), the differences can be summarized as follows.
\begin{itemize}
\item \emph{Depth.}  SM achieves the minimal depth, in fact BSGS generally requires one or two additional levels.
\item \emph{Noise.} In SM, the final variance grows with $a_j^2$ for the monomials of maximal depth, whereas in BSGS the coefficients are absorbed by modulus switching (\Cref{teo.bsgs}). This advantage, however, comes from applying the coefficients \emph{before} the last multiplication, and the same can be done in SM by scaling one of the two factors before computing the final product of each maximal-depth monomial. In this way, the coefficient contribution is absorbed by the subsequent modulus switching step, without increasing the multiplicative depth. 
\item \emph{Number of multiplications.} For a dense polynomial of degree $\kappa$, SM requires $\kappa-1$ ciphertext multiplications, and BSGS about $3\sqrt\kappa$. On the other hand, for a 
sparse polynomial, SM only computes the powers that are needed. 
\end{itemize}
Therefore, SM is preferable when minimizing the multiplicative depth, and hence the ciphertext modulus, is the main objective, as well as for sufficiently sparse polynomials. Conversely, BSGS is advantageous for large dense polynomials when reducing the number of ciphertext multiplications is the primary goal and one or two additional levels are acceptable.

\subsection{Rotation-Based Polynomial Evaluation}
Polynomial evaluation methods considered so far rely only on additions and multiplications. Recently, the authors of \cite{10.1007/978-981-99-8736-8_3} showed that, for BGV and BFV, the number of ciphertext-ciphertext multiplications required to evaluate a fixed polynomial can be reduced by exploiting the Galois structure of plaintext slots of high algebraic rank.

Intuitively, the main idea is to express a polynomial in terms of the algebraic norm or trace. To ease the explanation, we focus on the first, which is also the one implemented by the authors in the Microsoft SEAL library \cite{sealcrypto}. According to this method, under~\cite[Heuristic 4.2.]{10.1007/978-981-99-8736-8_3} it is possible to write a polynomial $f$ of degree smaller than the rank $d$ of the plaintext slots as
\begin{equation}
f(x)=\mathrm{N}(\theta-x)-x^d-g(x),
\end{equation}
for a suitable slot element $\theta$ and a polynomial $g$, which can be precomputed at the beginning. As explained in~\cite[Section 4.1]{10.1007/978-981-99-8736-8_3} computing $\mathrm{N}(\theta-x)$ consists in $\log_2(d)$ multiplications of a ciphertext by a rotation of itself. The critical quantity resulting from this computation is modeled by our work in \Cref{sec: automorphisms}, whereas the computations of $x^d$ and $g(x)$ are well studied by the methods proposed above. 

Another more recent work \cite{cryptoeprint:2026/1089} extends this approach to more general slot structures and to polynomials whose degree exceeds the slot rank. Although the corresponding algorithms are different, they are built from the same basic operations considered in our analysis: squaring, product of possibly dependent ciphertexts, product involving rotations and addition of possibly dependent ciphertexts. Consequently, once the multiplicative depth of the intermediate ciphertexts and the levels at which the final additions are performed are identified, the variance estimates follow from \Cref{sec:theoric_contribution}, in the same way as for the other polynomial evaluation strategies considered in this section.
  
This further highlights that the framework developed in this paper is not tied to any specific polynomial evaluation strategy. Rather, it provides a general methodology that can  cover not only classic polynomial evaluation methods but also more innovative techniques.

\section{Experimental Validation}\label{sec:experiments}

In this section, we validate our theoretical results by comparing the estimates of the variance obtained through our approach with empirical observations (\Cref{sec:var}). Experimental evaluations of the variance were carried out using the OpenFHE library \cite{OpenFHE}. Note that OpenFHE does not provide dedicated functions for computing powers of a ciphertext or for evaluating polynomial functions on ciphertexts. Therefore, these operations must be implemented manually using the \texttt{EvalMult} and \texttt{EvalAdd} functions. In our implementation, ciphertext powers are computed using both the square-and-multiply and the baby-step giant-step algorithms. 

We also provide our own selection of the moduli based on our variance estimation method (\Cref{sec:mod}). We compare our parameters with those employed in OpenFHE and HElib \cite{helib}, showing that our approach allows us to use significantly smaller moduli, leading to considerable efficiency gains without compromising security. Indeed, because our variance estimation never underestimates the actual variance, the decryption failure probability is negligible\footnote{Note that the decryption failure probability is never exactly zero, as average-case analysis intrinsically relies on probabilistic bounds. }.

The code used for the experiments is publicly available at\\
\url{https://github.com/nadirmur/openFHE}

\subsection{Variance estimation}\label{sec:var}

The experiments in this section were performed for different values of the ring dimension $n$ and the plaintext modulus is set to $t = 65537$. For each value of $n$, the experimental variance was estimated by sampling 10000 different ciphertexts. For simplicity, we report the base-2 logarithm of the variance. 
In all the theoretical estimates reported below, we incorporate the correction introduced in \Cref{rem:epsi}, setting $\alpha=1/100$ and $\epsilon=6\alpha$. In particular, contributions of the form $4nV_{\rm ms}^2$ and $2nV_{\rm ms}^2$ are replaced by $(4+\epsilon)nV_{\rm ms}^2$ and $(2+\epsilon)nV_{\rm ms}^2$, respectively.

Given a ciphertext $\bm{c}$, \Cref{tab:squaring,tab:c7} report our estimate of the variance of the first coefficient of the critical quantity associated with $\bm{c}^2$ and $\bm{c}^7$, which can be computed according to \Cref{teo.mul.c1c2,teo.mul.c2}, respectively.  \Cref{tab:rotation5,tab:rotationnm1} analyze the variance of the first coefficient of the critical quantity associated with the product of a ciphertext by a rotation of itself, according to \Cref{thm:autoMS}. Specifically, we consider $\sigma_5$ and $\sigma_{n-1}$ as instances. 
\begin{figure}[H]
    \centering

    \begin{subfigure}{0.48\textwidth}
        \centering
        \footnotesize
        \setlength{\tabcolsep}{8pt}
        \begin{tabular}{cccc}
            $n$ & $2^{13}$ & $2^{14}$ & $2^{15}$ \\
            \midrule
            \textit{exp} & 96.61 & 99.65 & 102.65 \\
            \textit{our} & 96.68 & 99.68 & 102.68
        \end{tabular}
        \caption{Variances for $\bm{c}^2$.}
        \label{tab:squaring}
    \end{subfigure}
    \hfill
    \begin{subfigure}{0.48\textwidth}
        \centering
        \footnotesize
        \setlength{\tabcolsep}{8pt}
        \begin{tabular}{cccc}
            $n$ & $2^{13}$ & $2^{14}$ & $2^{15}$ \\
            \midrule
            \textit{exp} & 95.64 & 98.66 & 101.49 \\
            \textit{our} & 96.68 & 99.68 & 102.68
        \end{tabular}
        \caption{Variances for $\bm{c}^7$.}
        \label{tab:c7}
    \end{subfigure}

    \vspace{0.5cm}

    \begin{subfigure}{0.48\textwidth}
        \centering
        \footnotesize
        \setlength{\tabcolsep}{8pt}
        \begin{tabular}{cccc}
            $n$ & $2^{13}$ & $2^{14}$ & $2^{15}$ \\
            \midrule
            \textit{exp} & 95.65 & 98.68 & 101.67 \\
            \textit{our} & 95.70 & 98.70 & 101.70
        \end{tabular}
        \caption{Variances for $\nu_{\sigma_{5}}$.}
        \label{tab:rotation5}
    \end{subfigure}
    \hfill
    \begin{subfigure}{0.48\textwidth}
        \centering
        \footnotesize
        \setlength{\tabcolsep}{8pt}
        \begin{tabular}{cccc}
            $n$ & $2^{13}$ & $2^{14}$ & $2^{15}$ \\
            \midrule
            \textit{exp} & 95.65 & 98.65 & 101.65 \\
            \textit{our} & 95.70 & 98.70 & 101.70
        \end{tabular}
        \caption{Variances for $\nu_{\sigma_{n-1}}$.}
        \label{tab:rotationnm1}
    \end{subfigure}

    \caption{Comparison between our theoretical estimates and the empirical ones.}
    \label{fig:var-com}
\end{figure}
In \Cref{fig:cheb}, given a ciphertext $\bm{c}$, we report our (\textit{our}) estimate of the variance of the first coefficient of the critical quantity associated with two polynomials:
$T_5(x)=16x^5-20x^3+5 x$ and $g(x)=x^8 + x^7 + x^6 + x^5,$ considering Square and Multiply ({SM}) and Baby-Step Giant-Step ({BSGS}) algorithms. Moreover, the tag ``exp'' shows the observed values from OpenFHE \cite{OpenFHE} library with $10000$ polynomial samples setting the parameters according to the required multiplicative depth, with $t = 65537$, $\chi_s  = \U_3$, $\chi_e = \mathcal{DG}(0, \sigma^2)$, with $\sigma = 3.19$ and security level at least 128-bit.

\begin{figure}[H]
    \centering
    \begin{subfigure}[t]{0.48\textwidth}
        \centering
        \footnotesize
        \setlength{\tabcolsep}{4pt}
        \begin{tabular}{lccc}
            $n$ & $2^{13}$ & $2^{14}$ & $2^{15}$ \\
            \toprule
            \textit{exp} (SM) & 103.66 & 106.69 & 109.66 \\
            \textit{our} (\Cref{teo.mul.ckpow.sm})
                & 104.68 & 107.68 & 110.68 \\
            \midrule
            \textit{exp} (BSGS) & 95.70 & 98.62 & 101.67 \\
            \textit{our} (\Cref{teo.bsgs})
                & 96.68 & 99.68  & 102.68  \\
        \end{tabular}
        \caption{Variances for $T_5(\bm{c})$}
        \label{tab:cheb-var}
    \end{subfigure}
    \hfill
    \begin{subfigure}[t]{0.48\textwidth}
        \centering
        \footnotesize
        \setlength{\tabcolsep}{4pt}
        \begin{tabular}{lccc}
            $n$ & $2^{13}$ & $2^{14}$ & $2^{15}$ \\
            \toprule
            & 97.98 & 101.02 & 104.00 \\
            & 98.68 & 101.68 & 104.68 \\
            \midrule
            & 95.64 & 98.65 & 101.70 \\
            & 96.68 & 99.68  & 102.68 \\
        \end{tabular}
        \caption{Variances for $g(\bm{c})$}
        \label{tab:g-var}
    \end{subfigure}
    \caption{Comparison between our theoretical estimates and the empirical variances.}
    \label{fig:cheb}
\end{figure}
Moreover, to highlight how the choice between the square-and-multiply and baby-step giant-step algorithms affects noise growth, we consider the following two polynomials $h_1(x)=10x^{20}+10x^{16}+10x^3$ and $h_2(x)=10x^{20}+10x^{15}.$
This specific example is also interesting since one might expect \(h_1(x)\) to produce more noise than \(h_2(x)\), since it contains more monomials and its highest-degree monomial other than \(x^{20}\) has a larger degree. However, as \Cref{tab.h} shows, this is not the case.

\begin{figure}[H]
    \centering
    \begin{subfigure}[t]{0.48\textwidth}
        \centering
        \footnotesize
        \setlength{\tabcolsep}{4pt}
        \begin{tabular}{lccc}
            $n$ & $2^{13}$ & $2^{14}$ & $2^{15}$ \\
            \toprule
            \textit{exp} (SM)   & 102.30 & 105.27 & 108.28\\
             \textit{our} (\Cref{teo.mul.ckpow.sm}) & 103.33 & 106.33 & 109.33\\
            \midrule
             \textit{exp} (BSGS) & 95.64 & 98.66 & 101.64 \\
            \textit{our} (\Cref{teo.bsgs}) & 96.68 & 99.68 & 102.68  \\
        \end{tabular}
        \caption{Variances for $h_1(\bm{c})$}
        \label{tab:cheb-var}
    \end{subfigure}
    \hfill
    \begin{subfigure}[t]{0.48\textwidth}
        \centering
        \footnotesize
        \setlength{\tabcolsep}{4pt}
        \begin{tabular}{lccc}
            $n$ & $2^{13}$ & $2^{14}$ & $2^{15}$ \\
            \toprule
            & 102.30 & 105.27 & 108.28\\
            & 103.33 & 106.33 & 109.33\\
            \midrule
            & 103.87 & 106.93 & 109.89 \\
            & 104.33 & 107.33 & 110.33\\
        \end{tabular}
        \caption{Variances for $h_2(\bm{c})$}
        \label{tab:h-var}
    \end{subfigure}

    \caption{Comparison between our theoretical and the empirical variances.}
    \label{tab.h}
\end{figure}

As shown in the tables, our method provides very accurate estimates of the variance with differences from the experimental values below $1.2$ bits. Moreover, this high level of accuracy is achieved without ever underestimating it, which is essential for ensuring both safety and correct system operation.

\subsection{Selection of Ciphertext Moduli}\label{sec:mod}

\Cref{sec:theoric_contribution} provides accurate estimates for the variance of the coefficients of the critical quantity after the main homomorphic operations. Combined with the fact that such coefficients can be accurately modeled by Gaussian distributions\footnote{We refer to \cite{cryptoeprint:2025/2027} for a detailed discussion on this assumption.}, these estimates allow us to derive tight probabilistic bounds on the error norm. Consequently, it is possible to determine suitable conditions on the ciphertext moduli to guarantee the correct functioning of the scheme. 

Indeed, if the coefficients of the critical quantity $\nu$ are modeled as Gaussian random variables with variance $V$, then the following bound on the decryption failure probability holds \begin{equation*}
\mathbb{P}\left( |\nu|_{\infty} > \frac{q}{2} \right)
\leq
n \left(1-\text{erf}\left(\frac{q}{2\sqrt{2V}}\right)\right).
\end{equation*}  
Hence, by defining a \textit{security parameter} \( D \) such that
$D \leq {q}/{2\sqrt{2V}}$, it follows that \begin{equation*}
\mathbb{P}\left( \|\nu\|_{\infty} > \frac{q}{2} \right) \leq n \left(1 - \text{erf}(D) \right),
\end{equation*}
which means that fixing the modulus  $q \ge 2D\sqrt{2V}$, we can properly select $D$ to guarantee that the probability of decryption failure remains negligibly small. A common choice for practical usage is $D=8$ since, for instance, it results in decryption failure probability of approximately $2^{-83}$ when $n=2^{13}$.

\paragraph{On the Gaussian Assumption.}
As widely discussed in \cite{cryptoeprint:2025/2027}, we can model the coefficients of the critical quantities involved in any circuit of operations as Gaussian thanks to the systematic usage of modulus switch. Specifically, the authors of~\cite{cryptoeprint:2025/2027} show that it is necessary to impose a lower bound on the ciphertext moduli ratio $p$, that is 
\begin{equation}\label{eq.p}
    p > \sqrt{\frac{V}{\alpha \Vms}}
\end{equation}
where $V$ denotes the variance before modulus switch is computed, $\Vms$ is as in \Cref{Vms} and $\alpha$ is a parameter which is heuristically selected as $\alpha= \sfrac{1}{100}$.
Note that the primes currently adopted by most existing libraries already satisfy this requirement. Hence, this does not compromise the modulus-size improvement proposed in this work compared to the state-of-the-art. 
\ec

\paragraph{Comparison with OpenFHE.}

Given a multiplicative depth $M$, OpenFHE uses $M+2$ moduli in order to allow an additional modulus switching after the final multiplication. In our method, the primes $p_0, \ldots, p_{M+1}$ whose product provides the ciphertext modulus $q$ are selected imposing the condition \eqref{eq.p}, namely, for $0<\ell<M+1$,
$p_\ell > \sqrt{{V}/{\alpha \Vms}}$,
 where $V$ is our variance estimate and $\alpha = 1/100$. The primes $p_0$ and $p_{M+1}$ must satisfy
$p_0 > 2 D \sqrt{2 \Vms},$ and  $p_{M+1} > \sqrt{{V_\mathsf{clean}}/{\alpha \Vms}},$
where $V_\mathsf{clean}$ is the variance of a fresh ciphertext. In the cases $n = 2^{13}, 2^{14}, 2^{15}$, the variance of a fresh ciphertext is $48.76, 49.76, 50.76$, respectively (see \cite{cryptoeprint:2025/2027}). For a fair comparison, each prime is chosen as the smallest NTT-friendly prime, satisfying the corresponding lower bound, with all primes distinct and different from the plaintext modulus. 

Observe that when selecting the moduli in this case, not only the multiplicative depth must be taken into account, but also the number of additions required to perform ciphertext-scalar multiplications. In OpenFHE, this parameter is specified through \texttt{SetEvalAddCount}. For example, in the evaluation of the Chebyshev polynomial, the required number of additions is equal to 16, while the multiplicative depth is 3.

In Figure~\ref{fig:moduli}, we compare the ciphertext modulus bit sizes selected by OpenFHE with those obtained using our variance estimation method when $\bm{c}^2$ and $\bm{c}^7$ are computed. 

\begin{figure}[H]
    \centering
    \begin{subfigure}{0.48\textwidth}
        \centering
        \footnotesize
        \setlength{\tabcolsep}{8pt}
        \begin{tabular}{cccc}
            $n$  &$2^{13}$ & $2^{14}$ & $2^{15}$ \\
            \midrule
            \textit{OpenFHE} & 82.8 & 85.3 & 88.6 \\
            \textit{Our method} & 73.0 & 75.0 & 78.8  \\
        \end{tabular}
        \caption{Modulus bit sizes for computing $\bm{c}^2$ (multiplicative depth 1).}
        \label{tab:q3}
    \end{subfigure}
    \hfill
    \begin{subfigure}{0.48\textwidth}
        \centering
        \footnotesize
        \setlength{\tabcolsep}{8pt}
        \begin{tabular}{cccc}
            $n$ & $2^{13}$ & $2^{14}$ & $2^{15}$ \\
            \midrule
            \textit{OpenFHE}    & 154.8 & 159.3 & 164.6 \\
            \textit{Our method} & 135.5 & 139.5 & 145.3 \\
        \end{tabular}
        \caption{Modulus bit sizes for computing $\bm{c}^7$ (multiplicative depth 3).}
        \label{tab:q5}
    \end{subfigure}
    \caption{Comparison of ciphertext modulus bit sizes between OpenFHE (v1.5.1) and our selection method.}
    \label{fig:moduli}
\end{figure}

In particular, Figure \ref{fig:var-com} shows that our estimates are very tight with respect to the actual variance of the critical quantity, while never underestimating it in any of our experiments. As a result, our approach enables a more efficient modulus selection, minimizing the modulus size without compromising correctness and security, as the probability of decryption failure remains negligible. Similar results can be observed in \Cref{fig:moduli2} for the moduli size for computing the polynomials $T_5(\bm{c})$ and $g(\bm{c})$, using the square and multiply method (\Cref{sec.SM}).

\begin{figure}[H]
    \centering
    \begin{subfigure}{0.48\textwidth}
        \centering
        \footnotesize
        \setlength{\tabcolsep}{8pt}
        \begin{tabular}{cccc}
            $n$  &$2^{13}$ & $2^{14}$ & $2^{15}$ \\
            \midrule
            \textit{OpenFHE}    & 160.8 & 165.3 & 170.6 \\
            \textit{Our method} & 143.8 & 147.8 & 153.6 \\
        \end{tabular}
        \caption{Modulus bit sizes for computing  $T_5(\bm{c})$ (multiplicative depth 3 and addition count 16).}
        \label{tab:cheb-mod2}
        \medskip
        \medskip
    \end{subfigure}
    \hfill
    \begin{subfigure}{0.48\textwidth}
        \centering
        \footnotesize
        \setlength{\tabcolsep}{8pt}
        \begin{tabular}{cccc}
            $n$ & $2^{13}$ & $2^{14}$ & $2^{15}$ \\
            \midrule
            \textit{OpenFHE}    & 154.8 & 159.3 & 164.6 \\
            \textit{Our method} & 136.5 & 140.5 & 146.3 \\
        \end{tabular}
        \caption{Modulus bit sizes for computing $g(\bm{c})$ (multiplicative depth 3 and addition count 3).}
        \label{tab:c8-mod}
        \medskip
        \medskip
    \end{subfigure}
    \caption{Comparison of ciphertext modulus bit sizes between OpenFHE (v1.5.1) and our selection method.}
    \label{fig:moduli2}
\end{figure}

\paragraph{Comparison with HElib.}
Here we provide a comparison of our moduli selection with respect to those currently employed in HElib \cite{halevi2020design}. Note that in \cite{costache2023optimisations} a strategy for optimizing HElib parameters is proposed. However, we do not provide a direct comparison with this approach, since it is tailored to the specific library and optimizes a different objective, namely the trade-off involving the special modulus used for key switching, which we deliberately do not model.  For reference, \cite[Table 8]{costache2023optimisations} results in a modulus ratio of $32/33/34$ bits (under the most permissive setting), while our framework gives $31/32/33$ bits, for $n={2^{13},2^{14},2^{15}}$. However, since both the optimization objectives and parameter settings differ, these results should not be interpreted as a direct comparison.  
Indeed, in this work, we intend to provide a general framework for optimizing the ciphertext moduli selection that is library independent. 

Unlike OpenFHE, the latest version of HElib does not select the moduli dimension based on the multiplicative depth of a circuit. Instead, the parameter selection is based on the bit length of the largest modulus in the chain. Hence, to have a significant comparison, we follow the same strategy of \cite{costache2023optimisations} and compare the ratio between the moduli of successive levels.
In HElib, by construction, this ratio is always larger than 36 bits. However, as pointed out in \cite{costache2023optimisations}, in practical usage it is always much larger, with size of approximately 54 bits on average, for ring dimensions $n \in \{2^{13}, 2^{14}, 2^{15}\}$. Instead, in our method, we only require that the primes $p_\ell$, for $0 < \ell < M+1$ ($M$ multiplicative depth), satisfy \cref{eq.p}, namely
$p_\ell > \sqrt{{V}/{\alpha \Vms}},$
where $V$ denotes the variance estimate obtained from our bounds prior to modulus switching and is determined by the structure of the evaluating circuit. As an illustrative example, we consider a circuit where, at each level, a multiplication between two dependent ciphertexts occurs. In this context, \Cref{teo.mul.c2} and \Cref{rem:epsi} yield the estimate $V \approx (4+\epsilon)n\Vms^2$, leading to the constraint $$p_\ell >  \sqrt{\frac{(4+\epsilon)n\Vms}{\alpha}}.$$ 
We use this scenario as a reference setting for the comparisons presented in \Cref{tab.ratioHElib}. Different circuits may yield distinct values of $V$, depending on the sequence of homomorphic operations performed. However, our analysis offers a systematic approach to accurately estimate the corresponding variance in each case and to derive the associated modulus constraints.

In \Cref{tab.ratioHElib} we present the bit-size of the ratio between adjacent moduli resulting from our approach, for $n \in \{2^{13},2^{14},2^{15} \}$. The experiments assume ternary secret keys with Hamming weight $h=n/2$, in accordance with the library specifications, and plaintext modulus $t=65537$. 
\begin{table}[h!]
\centering
\begin{tabular}{cccccccc}
 $n$  && $2^{13}$ && $ 2^{14}$&& $2^{15}$ \\
\midrule
$\log_2(p_\ell)$  && $31.04$ && $32.04$ && $33.04$ \\
\\
\end{tabular}
\caption{ Ratio between adjacent ciphertext moduli for different ring dimensions $n$, according to our approach, represented using their bit size.}
\label{tab.ratioHElib}
\end{table}

The results show that our approach consistently produces substantially smaller modulus ratios, achieving reductions of up to $5$ bits between consecutive ciphertext moduli compared to the smallest possible ratio dimension in HElib. The improvement becomes considerably more significant when compared with the modulus ratios typically adopted in practical deployments, where the reduction can reach $23$ bits per level. \ec

\section{Conclusions}\label{sec:conclusions}
This work provides the first average-case analysis for the BGV scheme that addresses the case of ciphertext dependencies. 
We show that such dependencies have a non-negligible impact on homomorphic computations, as they increase the expected variance of the error coefficients involved in operations between ciphertexts. 
At the same time, we show that these effects can be effectively controlled in practice, thanks to modulus switching, which helps attenuate the contribution of dependency-induced correlations.

We derive accurate variance estimates for homomorphic additions and multiplications involving dependent ciphertexts, accounting not just for secret/public-key-induced correlations but also for those induced by ciphertext itself. In particular, we extend our analysis to the homomorphic evaluation of polynomial functions, a central component of practical FHE applications. This allows us to obtain tight bounds on the resulting error growth.

Experimental findings indicate that the proposed estimates align closely with observed variances, with discrepancies between theoretical and experimental values never exceeding 1.2 bits and consistently avoiding underestimation.

Finally, these estimates are leveraged for ciphertext modulus selection, leading to significantly smaller parameters and improved efficiency compared to those currently adopted in state-of-the-art FHE libraries. Specifically, in comparison with OpenFHE, the methodology reduces the ciphertext modulus by an average of approximately $16.3$ bits across all evaluated benchmarks, corresponding to an average reduction of about $11.5\%$. Additionally, relative to the modulus ratios currently employed in HElib, the approach can decrease the size of the ratios between consecutive moduli by $23$ bits per level in practical parameter settings. 

Future work could investigate the applicability of the proposed framework to other RLWE-based FHE schemes. For example, we believe that 
for the BFV scheme, an adaptation may require a more complicated investigation, due to the different role played by the modulus switching technique. 

\section*{Acknowledgment}
The second and third authors were partially supported by the Italian Ministry of University and Research in the framework of the Call for Proposals for scrolling of final rankings of the PRIN 2022 call - Protocol no. 2022RFAZCJ.

\bibliographystyle{splncs04}
\bibliography{bibliography}

\appendix

\section{Proof of Lemma~\ref{lemma:surjective}}\label{appendix-lemma:surjective}

   We show by contradiction that if $\pi$ is not surjective, then $\dim_{\mathbb{F}_p}J\le n/2$.
    Suppose $\pi(J) \subsetneq \mathbb{F}_p^2$, hence it is a subspace of dimension less than two over $\mathbb{F}_p$, which means that there exists $\rho,\beta \in \mathbb{F}_p$ such that $(\rho,\beta)\ne\{(0,0)\}$  and   \begin{equation}\label{eq:linear-rel}
        \rho a|_{j_1} + \beta a|_{j_2} = 0 \quad \forall a \in J.
    \end{equation}
    We now show that there exists a binomial $b$ such that   \begin{equation}\label{eq:binomial}
        ab = 0 \quad \forall a \in J.
    \end{equation}
    Indeed, if we define $b \coloneqq \rho x^{-j_1} + \beta x^{-j_2} \in \R_p$, using the bilinearity of the product, \begin{equation}\label{eq:coeff-zero}
        ab|_0=\rho (ax^{-j_1})|_0 + \beta (ax^{-j_2})|_0 = \rho a|_{j_1} + \beta a|_{j_2}=0,
    \end{equation}
    where the last equality follows from \Cref{eq:linear-rel}.
    Since $J$ is an ideal, it is closed with respect to multiplication of any $x^i$, meaning that \Cref{eq:coeff-zero} holds for any coefficient $i$, which implies \Cref{eq:binomial}. 
     This means that $x^n+1 \mid ab$ for all $a\in J$. Hence,
\begin{equation*}
    J\subseteq \operatorname{Ann}(b),
\end{equation*}
where $\operatorname{Ann}(b) \coloneqq \{a\in\mathcal{R}_p : ab=0\}$ denotes the annihilator of $b$.  Since
$x^n=-1$ in $\mathcal{R}_p$, multiplication by any power of $x$ is
invertible. Therefore, multiplying $b$ by a suitable power of $x$ does
not change the dimension of its annihilator. Let
\begin{equation*}
r\coloneqq \min\{|j_1-j_2|,\,n-|j_1-j_2|\}.
\end{equation*}
Since $j_1\neq j_2$, we have $1\leq r\leq n/2$. Using $x^n=-1$, after
multiplying $b$ by a suitable power of $x$, we can write
\begin{equation*}
b'=\rho' + \beta' x^r, \qquad \rho',\beta'\in\mathbb{F}_p^*.
\end{equation*} 
Defining $g(x)=\gcd(x^n+1,b'(x))$, it is possible to verify that
\begin{equation*}
\dim_{\mathbb{F}_p}\operatorname{Ann}(b')=\deg(g).
\end{equation*}
Moreover, since $g$ divides $b'$, $\deg(g)\leq \deg(b')=r\leq \frac{n}{2}$, hence,
\begin{equation*}
\dim_{\mathbb{F}_p}J \leq \dim_{\mathbb{F}_p}\operatorname{Ann}(b) = \dim_{\mathbb{F}_p}\operatorname{Ann}(b') \leq \frac{n}{2},
\end{equation*}
which contradicts the assumption $\dim_{\mathbb{F}_p}J>\frac{n}{2}$, concluding the proof.

\section{Proof of \Cref{teo.bsgs}}\label{proof.teo.bsgs}
Throughout the proof, for any ciphertext $\bm{c}$, we denote by $\nu_{\bm c}$ its critical quantity.
According to \Cref{eq:bsgs}, we write $f(x)$ as
$$
f(x)=\sum_{j=0}^{\rho}x^{jm}B_j(x), \mbox{ where } B_j(x)=a_{jm}+f_j(x) \mbox{ and } f_j(x)=\sum_{k=1}^{m-1}a_{jm+k}x^{k},
$$
with $a_l=0$ for $l>\kappa$. We start by defining the multiplication levels at which powers of the ciphertexts involved in the evaluation are computed. For $1\le j\le\rho$, we define $\ell_{j}:=\lceil\log_2 m\rceil+\lceil\log_2 j\rceil$. Then,
\begin{itemize}
    \item any baby-step power $\bm c^{\,k}$, with $1\le k\le m$, lies at level 
    $\lceil\log_2 k\rceil\le\lceil\log_2 m\rceil$. In particular, $\bm c^{\,m}$ lies at level 
    $\ell_1=\lceil\log_2 m\rceil$;
    \item any giant-step power $(\bm c^{\,m})^{j}=\bm c^{\,jm}$, with $2\le j\le\rho$, 
    lies at level $\ell_j$, since $\bm c^{\,m}$ corresponds to level $\lceil\log_2 m\rceil$.
\end{itemize}
\noindent
Each block $B_j(\bm c)$ contains at most $m$ monomials,
and hence at most $m-1$ additions. Specifically, the
last block contains $\beta+1$ monomials, while the
other blocks contain at most $m$ monomials. We start by distinguishing two cases: $\rho=1$ and $\rho\ge 2$.
\paragraph{Case $\rho=1$.}
Since $\rho=\lfloor\kappa/m\rfloor$, then $\kappa<2m$. 
Moreover, $m=\lceil\sqrt{\kappa+1}\,\rceil$ gives $(m-1)^2<\kappa+1<2m+1$, that is, $m<4$. Since 
$\kappa\ge2$, then $m\in\{2,3\}$.\\
Now, we again distinguish between two cases: $\beta=0$ and $\beta\neq0$. Let us start by supposing that $\beta=0$. In this case,  we have $\kappa=m$. Since $m\in\{2,3\}$, we must have $m=2$: indeed, if $m=3$, we would have
$m=\lceil\sqrt{\kappa+1}\rceil=\lceil\sqrt{4}\rceil=2$,
a contradiction. Therefore, $\kappa=m=2$ and
\[
f(x)=a_1x+a_2x^2.
\]
In particular, $f_1(x)=0$ and
$B_1(\bm c)=a_2$, so the block reduces to $a_2\bm c^2$, which lies at level $\ell_1=1$. Since $\bm c^2$ is a
square, \Cref{teo.mul.c2} gives
\[
\V(\nu_{a_2\bm c^2}|_i)
\approx 4na_2^2\Vms^2.
\]
Moreover, $B_0(\bm c)=a_1\bm c$ is modulus-switched
to level $\ell_1$ before the addition, contributing
$\Vms$. Thus,
\[
\V(\nu_f|_i)
\le \Vms+4na_2^2\Vms^2,
\]
thus concluding this case.
Instead, suppose now that $\beta\ne0$, then $f_1(x)\ne0$, since $a_\kappa\ne0$ is the coefficient of $x^\beta$ in 
$f_1(x)$. Thus:
\begin{itemize}
    \item For $B_1$, the ciphertext $\bm c^{\,m}$ lies at level $\ell_1=\lceil\log_2 m\rceil$, 
    while $\bm c_{f_1}$ lies at level at most $\lceil\log_2(m-1)\rceil$. Since $m\in\{2,3\}$, we 
    have $\lceil\log_2(m-1)\rceil<\lceil\log_2 m\rceil$, so the modulus switching procedure is 
    applied to $\bm c_{f_1}$ to obtain the same modulus as $\bm c^{\,m}$, and 
    $\V(\nu_{\bm c_{f_1}}|_i)=\Vms$. By \Cref{teo.mul.c1c2}, 
    $\V(\nu_{\bm c^{\,m}\bm c_{f_1}}|_i)\approx 4n\Vms^2$, at level $\ell_1+1$. The summand 
    $a_m\bm c^{\,m}$ lies at level $\ell_1$ and is modulus-switched before the addition, hence 
    $\V(\nu_{\bm c^{\,m}B_1(\bm c)}|_i)\approx\Vms+ 4n\Vms^2$.
    \item The block $B_0$ contains the baby-step powers $\bm c^{\,k}$, $k\le m-1$, and hence lies 
    at level at most $\lceil\log_2(m-1)\rceil<\ell_1+1$.
\end{itemize}
Thus, in order to sum the two blocks, the modulus switching procedure is applied to $B_0(\bm c)$ 
and, by \Cref{prop.add.dep}, $\V(\nu_f|_i)\le2\Vms+  4n\Vms^2$, concluding the case $\rho =1$.

\paragraph{Case $\rho \ge 2$.} We now analyze the blocks $B_j(\bm c)$ separately, by considering the cases $ j \in \{2,\dots,\rho\}$, $j=1$ and $j=0$.

\begin{itemize}
\item  $B_j(x)$, where $2\le j\leq\rho$. If $f_j(x)\ne 0$, all the ciphertext powers appearing in 
$f_j(x)$ lie at level at most $\lceil\log_2(m-1)\rceil\le\lceil\log_2 m\rceil < \ell_j$, so that 
$\bm c_{f_j}$ lies at a level smaller than $\ell_j$. Hence, in order to multiply $\bm c_{f_j}$ by 
$\bm c^{\,jm}$, modulus switch must be applied to $\bm c_{f_j}$ to obtain the same modulus as $\bm c^{\,jm}$. Thus $\V(\nu_{\bm c_{f_j}}|_i)=\Vms$.
Now, to compute $\bm c^{\,jm}B_j(\bm c)$, 
we apply \Cref{teo.mul.c1c2} to $\bm c^{\,jm}\bm c_{f_j}$ and we obtain $\V(\nu_{\bm c^{\,jm}\bm c_{f_j}} |_i)\approx  4  n\Vms^{2}$ (which lies at level  $\ell_{j}+1$) and,
since the remaining summand 
$a_{jm}\bm c^{\,jm}$ lies at level $\ell_{j}$, the modulus-switch is applied before the addition.
Hence the block $\bm c^{\,jm}B_j(\bm c)$ satisfies
\[
  \V(\nu_{\bm c^{\,jm}B_j(\bm c)}|_i)\;\approx \Vms + 4 n\Vms^{2} ,\qquad\text{at level } \ell_j+1, 
\]

If, instead, $f_j(x)=0$, the block reduces to the monomial $a_{jm}\bm c^{\,jm}$.
Thus, by \Cref{teo.mul.c2,teo.mul.c1c2},
\[
  \V(\nu_{\bm c^{\,jm}B_j(\bm c)}|_i)\approx 4 \,n\,a_{jm}^{2}\Vms^{2},\qquad\text{at level }\ell_j,
\]
Note that if $\beta\ne 0$, then $f_\rho(x)\ne 0$, and the maximum level is $d=\ell_\rho+1$. Otherwise, if $\beta=0$, then $f_\rho(x)=0$, and
$B_\rho\ne0$, so $\bm c^{\,\rho m}B_\rho(\bm c)= a_\kappa\bm c^{\,\rho m}$ is nonzero and lies at level $\ell_\rho$.
\item For the block $B_1(x)$, $\bm c^{\,m}$ lies at level $\ell_1=\lceil\log_2 m\rceil$. If 
$f_1(x)=0$, the block reduces to $a_m\bm c^{\,m}$ and lies at level $\ell_1$. If $f_1(x)\ne0$ and 
$\lceil\log_2(m-1)\rceil<\lceil\log_2 m\rceil$, then $\bm c_{f_1}$ lies at a level smaller than 
$\ell_1$ and the same computation as in the previous item applies. Otherwise $\bm c_{f_1}$ may 
already lie at level $\ell_1$, and no modulus switching is applied before the product. In both 
cases $\bm c^{\,m}B_1(\bm c)$ lies at level $\ell_1+1$.

Now, the maximal level satisfies 
$d\ge\ell_\rho$, with $d=\ell_\rho+1$ if $\beta\ge1$. Hence $\bm c^{\,m}B_1(\bm c)$ lies at a 
level smaller than $d$ whenever $\rho\ge3$ (since $\ell_1+1<\ell_\rho$) or $\rho=2$ and 
$\beta\ge1$ (because $\ell_1+1=\ell_2<d$). In these cases the modulus switching procedure is applied 
before the final addition and its contribution is $\Vms$.

In the remaining case, $\beta=0$ and $\rho=2$ (and $m=3$). So, if $f_1(x)\ne0$ the first case applies since $\ell_1+1=\ell_2=d$ and
$\V(\nu_{\bm c^{\,m}B_1(\bm c)}|_i)\approx\Vms+ 4 n\Vms^{2}$. If 
$f_1(x)=0$, the block lies at level $\ell_1<d$ and contributes $\Vms$.
\item The first block $B_0(x)$ requires no giant multiplication, and hence lies at a level smaller than $d$.
\end{itemize}
\noindent 
Finally, in order to sum all the terms of $f(\bm c)$ together, all $\bm c^{\,jm}B_j(\bm c)$ must be at the same level  $d$. In particular, 
if $\beta\ge 1$,
\begin{itemize}
    \item for any $j$ such that $\lceil\log_2 j\rceil=\lceil\log_2\rho\rceil$ and $f_j(x)\ne0$, the block $\bm c^{\,jm}B_j(\bm c)$ already lies at level $d=\ell_\rho+1$. Instead, if $f_j(x)=0$, the block lies at level $\ell_\rho<d$ and is modulus-switched, contributing $\Vms$. Hence, let 
    $\hat{\bm c}=\sum_{\lceil\log_2 j\rceil=\lceil\log_2\rho\rceil}\bm c^{\,jm}B_j(\bm c)$. Thus, by \Cref{prop.add.dep}, 
    \begin{align*}
        \V(\nu_{\hat{\bm c}}|_i) \; & \approx \; \#\{0< j \leq\rho \,: \lceil\log_2 j\rceil=\lceil\log_2\rho\rceil\}\bigl(4n\Vms^{2}+\Vms\bigr)\\
        & = \#\{2^{\lceil\log_2\rho\rceil-1}< j \leq \rho\}\bigl(4n\Vms^{2}+\Vms\bigr)\\
         & = (\rho-2^{\lceil\log_2\rho\rceil-1})\bigl(4n\Vms^{2}+\Vms\bigr) = (\rho-A)\bigl(4 n\Vms^{2}+\Vms\bigr).
    \end{align*} 
    \item for any $j$ such that $\lceil\log_2 j\rceil<\lceil\log_2\rho\rceil$, and for $j=0$, we apply the modulus switching technique to $\bm c^{\,jm}B_j(\bm c)$. These blocks are (at most) $\rho+1-(\rho-A)$, since some blocks $B_j$ may be zero.  
\end{itemize}
Putting all together and applying \Cref{prop.add.dep}, we obtain,
\[
  \V(\nu_{f}|_i)\;\le\;(1+A)\Vms+(\rho-A)\bigl(4n\Vms^{2}+\Vms\bigr)
  \;=\;(\rho+1)\Vms+ 4 n(\rho-A)\Vms^{2}.
\]

\noindent If $\beta=0$, note that $\ell_\rho=\ell_{\rho-1}+1$, i.e. 
$\lceil\log_2\rho\rceil>\lceil\log_2(\rho-1)\rceil$, if and only if $\rho-1$ is a power of 
$2$. Note that if $f_j\neq 0$ for at least one 
$j\le\rho-1$ with $\lceil\log_2 j\rceil=\lceil\log_2(\rho-1)\rceil$, then the maximal level is 
$d=\ell_{\rho-1}+1$. Now,

\begin{itemize}
    \item if $\rho-1$ is not a power of $2$, then $\ell_{\rho-1}=\ell_\rho$, and we have two cases:
    \begin{itemize}
        \item if $f_{j}(x)\ne 0$, for at least a $j\leq \rho-1$ such that $\lceil\log_2 j\rceil=\lceil\log_2(\rho-1)\rceil=\lceil\log_2 \rho\rceil$, then $d=\ell_\rho+1$ and the  block $\bm c^{\,\rho m}B_\rho(\bm c)$ lies at level $\ell_\rho < d$ and is modulus-switched. The blocks lying at level $d$ are those with $1\leq \lceil\log_2 j\rceil=\lceil\log_2\rho\rceil$ and $j\le\rho-1$, which are $\rho-1-2^{\lceil\log_2\rho\rceil-1}=\rho-A-1$. Since the total number of  blocks is (at most) $\rho+1$, we have
        \[
          \V(\nu_{f}|_i)\le(\rho+1)\Vms+4n\bigl(\rho-1-2^{\lceil\log_2\rho\rceil-1}\bigr)\Vms^2 .
        \]
        \item if $f_j(x)=0$ for every $j\le\rho-1$ such that 
        $\lceil\log_2 j\rceil=\lceil\log_2(\rho-1)\rceil=\lceil\log_2\rho\rceil$, then $d=\ell_\rho$. The blocks $\bm c^{jm}B_j(\bm c)$ lying at level $d$ are the ones with
        \begin{itemize}
            \item $A<j\le\rho$, which are reduced to $a_{jm}\bm c^{\,jm}$ and each of them contributes 
            $4 n\,a_{jm}^{2}\Vms^{2}$;
            \item $\lceil\log_2 j\rceil=\lceil\log_2\rho\rceil-1$, i.e. 
            $2^{\lceil\log_2\rho\rceil-2}<j\le2^{\lceil\log_2\rho\rceil-1}$, which are at most 
            $A/2=2^{\lceil\log_2\rho\rceil-2}$ and their contribution is $\Vms + 4 n\Vms^{2}$ each. 
            \end{itemize}
            All the remaining blocks, which are at most 
            $\rho-A/2-(\rho-A-1)=1+A/2$, are modulus-switched, thus
            \[
              \V(\nu_{f}|_i)\le\bigl(1+A\bigr)\Vms+n\Bigl(2A+\sum_{A<j\le\rho} 4\,a_{jm}^{2}\Bigr)\Vms^{2},
            \]
        \end{itemize}
    \item if $\rho-1\geq 2$ is a power of $2$, then no $j<\rho$ satisfies 
$\lceil\log_2 j\rceil=\lceil\log_2\rho\rceil$, hence $d=\ell_\rho$. The blocks lying at level 
$d$ are $a_\kappa\bm c^{\,\rho m}$ and those with $\frac{\rho-1}{2}<j\le\rho-1$ and 
$f_j(x)\ne0$, which are at most $\frac{\rho-1}{2}$. The blocks in this range with $f_j(x)=0$ lie 
at level $\ell_j<d$ and are modulus-switched. 
Hence,
\[
  \V(\nu_{f}|_i)\le\rho\,\Vms+2n(\rho-1)\Vms^2+4 n\,a_\kappa^2\Vms^2 .
\]
    \item if $\rho=2$, then  
    thanks to the contribution of $B_0$ and $B_1$, we have
    \[
      \V(\nu_{f}|_i)\le 2\Vms+ 4 n\Vms^{2}+4n\,a_\kappa^2\Vms^2 .
        \]
\end{itemize}

\end{document}